\documentclass[final]{elsart}

\usepackage{amsmath}
\usepackage{amssymb,bm}
\usepackage{graphicx}
\usepackage{color}
\usepackage{mathrsfs} 
\usepackage{diagbox}
\usepackage{multirow}
\makeatletter
\newif\if@restonecol
\makeatother

\usepackage{algorithm}
\usepackage{algpseudocode}
\DeclareMathAlphabet\mathpzc{OT1}{pzc}{m}{it}
\let\mathcal=\mathpzc

\let\trueiiint=\iiint
\def\iiint{\mathop{\textstyle\trueiiint}\limits}
\def\intinfty{\int\limits_{\!\!-\infty\,\,}^{\,\,\infty\!\!}\kern-0.0em}
\def\iintinfty{\mathop{\int\!\!\int}\limits_{\!\!-\infty\,\,}^{\,\,\infty\!\!}\kern-0.0em}
\def\iiintinfty{\mathop{\int\!\!\int\!\!\int}\limits_{\!\!-\infty\,\,}^{\,\,\infty\!\!}\kern-0.0em}

\let\<=\langle
\let\>=\rangle
\let\^=\hat
\def\~#1{{\mbox{\sf#1}}}
\let\-=\mathbf

\def\circ{\ifmmode\mathchar"220E\else$\mathchar"220E$\fi}
\def\@#1{{\cal #1}}
\newtheorem{theorem}{Theorem}[section]

\newtheorem{proof}[theorem]{proof}

\numberwithin{equation}{section}

\journal{Elsevier}

\begin{document}
\centerline{}
\begin{frontmatter}



\title{Clustered active-subspace based local Gaussian Process emulator for high-dimensional and complex computer models}



\author[authorlabel1]{Junda Xiong}
\ead{jdxiong@sjtu.edu.cn}
\address[authorlabel1]{School of Mathematical Sciences,
          Shanghai Jiao Tong University, 800 Dongchuan Rd, 
          Shanghai 200240, P.R. China.}

\author[authorlabel2]{Xin Cai}
\ead{izumi\_xin@sjtu.edu.cn}
\address[authorlabel2]{School of Mathematical Sciences,
          Shanghai Jiao Tong University, 800 Dongchuan Rd, 
          Shanghai 200240, P.R. China.}

\author[authorlabel3]{Jinglai Li}
\ead{j.li.10@bham.ac.uk}
\address[authorlabel3]{School of Mathematics, University of Birmingham, Birmingham B15 2TT, UK. (Corresponding author)}

\medskip
\begin{center}
\end{center}
\begin{abstract}
Quantifying uncertainties in physical or engineering systems often requires a large number of simulations of the underlying computer models
that are computationally intensive. Emulators or surrogate models are often used to accelerate the computation in such problems, 
and in this regard the Gaussian Process (GP) emulator is a popular choice for its ability to quantify the approximation error in the emulator itself.
However, a major limitation of the GP emulator is that it can not handle problems of very high dimensions, 
which is often addressed with dimension reduction techniques.
In this work we hope to address 
an issue that the models of interest are so complex that they admit different low dimensional structures in different parameter regimes.
Building upon the active subspace method for dimension reduction, we propose a clustered active subspace method which identifies 
the local low-dimensional structures as well as the parameter regimes they are in (represented as clusters), and then construct
low dimensional and local GP emulators within the clusters. Specifically we design a clustering method based on 
the gradient information to identify these clusters, and a local GP construction procedure to construct the GP emulator within a local cluster. 
With numerical examples, we demonstrate that the proposed method is effective when the underlying models are of complex 
low-dimensional structures. 
\end{abstract}


\end{frontmatter}

\section{Introduction}
\label{sec:intro}

Computer models or simulators are one of the most important tools to study complex physical or biological processes and
explore their behaviors, in many fields of science and engineering. 
As is well known, the computer models are inevitably subject to various sources of uncertainty: imprecise boundary or initial conditions,
unknown model parameters, random perturbations, and so on.
To characterize
the impact of the uncertainties in the simulation results, a large number, e.g., tens of thousands or more, of simulations are required. 
On the other hand, in reality,
the computer models are often computationally intensive, especially when the system of interest involves highly complex physical processes, 
and in this case, conducting a large number of simulations to quantify the impact of uncertainties becomes a prohibitive task. 
A commonly used approach to overcome this
difficulty is to construct computationally inexpensive surrogate models, namely emulators, to approximate the computer
simulators in a probabilistic way \cite{sacks1989design}. Substantial efforts have been devoted to this topic and many different types of surrogate models 
have been developed:  the polynomial chaos expansion \cite{xiu2002wiener,xiu2007efficient}, radial basis functions (RBF) \cite{gutmann2001radial,regis2005constrained},  adaptive sparse grid collocation
\cite{ma2009adaptive}, the Gaussian processes (GP) regression~\cite{williams2006gaussian},  
and more recently, neural networks~\cite{zhang2019quantifying,qin2019data,zhu2019physics}. 

In this work, we choose to use the GP regression because of its probabilistic formulation, and we can take the advantage of 
the GP method's ability to epistemically quantify the uncertainty induced by all the random effects and limited number of simulations in a natural Bayesian framework~\cite{williams2006gaussian}. In particular, it makes it possible to statistically quantify the approximation error in the emulator result itself, 
which then can be incorporated in the analysis of the total uncertainty, see \cite{o1998uncertainty,oakley2002bayesian,oakley2004probabilistic}.
For this reason, the GP emulators have been widely used in the field of uncertainty quantification (UQ):
uncertainty propagation~\cite{bilionis2012multi,gorodetsky2016mercer,yang2020bifidelity}, parameter estimation~\cite{kandasamy2015bayesian,wang2018adaptive}, reliability analysis~\cite{bect2012sequential,wang2016gaussian,wu2016surrogate}, 
just to name a few.

Despite the wide success of the GP emulator, a major limitation of the method, as well as practically any aformentioned surrogate models, is
that it can not handle problems with high stochastic dimensionality.  Within the context of the GP model, the incapability  is due to the fact that 
the method relies on the Euclidean distance to define input-space correlations.
Since the Euclidean distance becomes uninformative as the dimensionality of the input space increases \cite{bengio2006curse}, 
unless the number of data points  available grows exponentially, an issue known as the curse of dimensionality. That is, directly reconstructing or approximating a function in a high dimensional space is not computationally feasible. 
To this end, considerable research efforts are focused on methodologies that can identify and exploit some special structure of the underlying mathematical model, and in particular some low dimensional structures. In such methods, it is assumed that the output of the model only depends on the inputs through its projection in a low dimensional subspace.
One first
 identifies such a low-dimensional subspace  and then construct the GP emulator in the obtained low dimensional subspace, where examples of this strategy include \cite{djolonga2013high,liu2017dimension,tripathy2016gaussian}.
In this regard a variety of dimension reduction techniques can be used to identify the low dimensional subspace, such as
the sliced inverse regression (SIR)~\cite{li1991sliced}, sliced average variance estimation~(SAVE), among many others (see, e.g., \cite{Bing2018Sufficient}).
Most of these dimension reduction methods only utilize the input-output data points to identify the low dimensional subspace. 
On the other hand, in many practical computer models, often the gradient of the output with respect to the input parameters is also available; for example, if the computer model is 
solved with a finite element method, the gradient can be readily computed using the adjoint method~\cite{biegler2003large}. 
When the model gradients are available, such information can also be used to identify the low dimensional structure. 
To this end, a gradient based dimension reduction approach, the so-called active subspace (AS) method,
has been developed~\cite{russi2010uncertainty,constantine2014active,constantine2015active}. 
Loosely speaking, the method identifies a low dimensional subspace of the original input space, termed as the active subspace, 
by considering the expectation of the gradient outer product~\cite{constantine2015active}.
Further details and theoretical analysis of the AS method are provided in Section~\ref{sec:probset}.

It is important to note here that  all  the aforementioned methods, with or without using gradient, are all based on the assumption that a global low dimensional structure of the underlying simulation model exists. In other word, it is assumed that the same low dimensional structure of the simulation model exists 
across the state space of the input parameter. 
As a result, the emulator is constructed within this global low dimensional subspace. 
This assumption, however, is not always true in practice, as many computer models are highly complex and may exhibit substantially different 
behaviors in different parameter regimes. 
In this case, assuming a global low dimensional structure of the underlying model becomes inappropriate. 
Rather, it is more flexible and sensible to assume that, should such low dimensional structures exist, they may only be valid locally, i,e., in a subregion of the entire space. 
Based on this idea, we propose a local dimensionality reduction scheme, which we call clustered active subspace method~(CAS). 
The main idea of the method is to partition the original input space into a number of subdomains, where each subdomain has its
own low dimensional structure. We proposed to identify such subdomains using a clustering method based on the data points.
Finally, we construct a local and low dimensional GP emulator of the underlying simulation model in each subdomain,
and as a result, we obtain a set of local GP emulators to mimic the behaviors of the underlying model.  

The rest of the paper is organized as follows. In Section \ref{sec:probset}, we describe the problem and give reviews to the main technical ingredients of our methods. Then in Section~\ref{sec:method} we present the clustered active subspace method and the construction of local GP based on it.
In Section \ref{sec:numeres}, we provide numerical results for three examples: two mathematical functions with known low dimensional structures
and a PDE based model. Finally Section \ref{sec:conclu} offers some closing remarks. 
\section{Technical preliminaries }\label{sec:probset}

\subsection{Problem setup}
We consider in this work a computational expensive function $y=f(\-{\bf x})$, where $\-{\bf x}$ is a $d$ dimensional random variable with distribution 
$\pi(\-{\bf x})$ and $y$ is a scalar. Assume that the support of $\pi(\-{\bf x})$, $\Omega$ is a compact subspace of $R^d$.  
In practice, the function $f(\-{\bf x})$ can be described by a complex PDE or  a large scale ODE system. 
When solving such systems, in addition to  evaluating the function values, the gradient of the function can often be obtained 
as a byproduct. We assume that this is the case for problems considered here.
Suppose that we have a set of data points 
\begin{equation}
D=\{({\bf x}^{(n)},y^{(n)})|y^{(n)}=f(\-{\bf x}^{(n)})\}_{n=1}^N,\label{e:tset}
\end{equation}
 with ${\bf x}^{(n)}$  drawn according to $\pi(\-{\bf x})$,
and  our goal here is to construct a surrogate model for $f(\-{\bf x})$ which can be used in other computational tasks. 
More precisely, we plan to use the GP regression to construct the surrogate, and the method is described in Section~\ref{sec:gp}. 
A main challenge here is that in many practical problems the dimensionality $d$ is large (for example $d\geq 100$), and  constructing GP model or any other surrogates  in such a high-dimensional setting is extremely difficult. 
As has been discussed in Section~\ref{sec:intro}, we will first reduce the dimensionality of the problem before constructing the GP model. 

\subsection{Gaussian Process}\label{sec:gp}
The GP regression performs a nonparametric regression in a Bayesian framework~\cite{williams2006gaussian}.
The basic idea of the GP method is to assume that the function of interest $f({\bf x},\epsilon)$ is a realization from a Gaussian random field,
whose mean is $\mu({\bf x})$ and covariance is specified by
a kernel function $k({\bf x},{\bf x}')$, namely,
\[ \mathrm{Cov}[f({\bf x}),f({\bf x}')] = k({\bf x},{\bf x}'). \]
The kernel $k({\bf x},{\bf x}')$ is positive semidefinite and bounded.
Commonly used kernel functions include the squared exponential kernel, the Matern kernel
and the rational quadratic kernel. 

Now given the data points $\{({\bf x}^{(n)},y^{(n)})\}_{n=1}^N$,
we want to predict the value of $y$ at a new point $\-{\bf x}$.
Now we let $\-X := \left[{\bf x}^{(1)}; \ldots; {\bf x}^{(N)}\right]$,
and $\-Y =[y^{(1)},\ldots, y^{(N)}]^{\mathrm{T}}$.
Under the GP assumption,  it is easy to see that the joint distribution of $(\-Y,\,y)$ is Gaussian,
\begin{equation}
\left[ \begin{array}{c}
         \-Y \\
        y \end{array} \right] \sim \mathcal{N}\left(\begin{array}{c}
         \mu(\-X) \\
        \mu(\-{\bf x}) \end{array},
        \left[
        \begin{array}{ll}
         K(\-X,\-X)+\sigma_M^2I &K(\-X,{\bf x}) \\
        K({\bf x},\-X) &K({\bf x},{\bf x}) \end{array}\right]
        \right) , \label{e:jointdis}
\end{equation}
where $\sigma_M^2$ is the variance of observation noise, $I$ is
an identity matrix, and the notation $K(\-A,\-B)$ denotes the matrix of the covariance evaluated at all pairs of points in set $\-A$ and in set $\-B$ using
the kernel function $k(\cdot,\cdot)$.
It follows immediately from Eq.~\eqref{e:jointdis} that  the conditional distribution $\pi_{GP}(y|\-{\bf x},\-X,\-Y)$  is also Gaussian:
\begin{subequations}
\label{e:gp}
  \begin{equation}
  \pi_{GP}(y|{\bf x},\-X,\-Y) =\mathcal{N}(\mu_\mathrm{pos}, \sigma^2_\mathrm{pos}),
  \end{equation}
where the posterior mean and variance are,
\begin{align}
&\mu_\mathrm{pos}({\bf x})=\mu({\bf x})+k({\bf x},\-X)(k(\-X,\-X)+\sigma_M^2I)^{-1}(\-Y-\mu({\bf x})),\\
&\sigma^2_\mathrm{pos}({\bf x}) = k({\bf x},{\bf x})-k({\bf x},\-X)(k(\-X,\-X)+\sigma_M^2I)^{-1}k(\-X,{\bf x}).
\end{align}
\end{subequations}
There are also a number of technical issues in the GP model, such as choosing the kernel function and determining the hyperparameters.
For detailed discussion of these matters, we refer the readers to~\cite{williams2006gaussian}.
As is well known, the GP method fails when the dimensionality of data is extremely high. It is one of the key issues to be addressed when applying GP based methods to high dimensional  problems.

\subsection{Active Subspace method}\label{sec:as}
We now discuss the dimension reduction technique that will be used in this work. 
Recall that it is assumed that the gradient of the target function is available in the problems under consideration. Thus we  use the so-called active subspapce (AS) method, which identifies the low dimensional subspace by using the gradient information. 
In what follows we give a brief introduction to AS, largely following \cite{constantine2014active}.

Consider  a target function $y=f(\-{\bf x})$ with $\-{\bf x}\in \mathbb{R}^d$ that is absolutely continuous and square-integrable with respect to a probability density function $\pi:\mathbb{R}^{d}\to\mathbb{R}_{+}$.
Now recall that we can compute the gradient of $f$ denoted by the column vector $\nabla_{{\bf x}}f({\bf x})=\left[\frac{\partial f}{\partial x_{1}}\,\,\cdots\,\,\frac{\partial f}{\partial x_{d}}\right]^{\mathrm{T}}$. 
Next we shall define the $d\times d$ matrix ${\bf C}$ as
\begin{equation}\label{asm1}
  {\bf C}=\mathbb{E}_{\pi({\bf x})}[(\nabla_{{\bf x}}f)(\nabla_{{\bf x}}f)^{\mathrm{T}}]
\end{equation}
where we assume that the products partial derivatives are integrable. Since ${\bf C}$ is symmetric positive definite, it can be decomposed as, 
\begin{equation}
  {\bf C = V\Lambda V}^{\mathrm{T}}
\end{equation}
where ${\bf \Lambda}=diag(\lambda_{1},\cdots,\lambda_{d})$ is a diagonal matrix with the eigenvalues of {\bf C} in decreasing order, $\lambda_{1}\geq\cdots\geq\lambda_{d}\geq 0$, and ${\bf V}\in\mathbb{R}^{d\times d}$ an orthonormal matrix consists of eigenvectors of ${\bf C}$.

Assuming that the reduced dimensionality is $r$,  we can partition the eigenvalues and eigenvectors into two parts:
\begin{equation}
  {\bf \Lambda} = \left[
  \begin{matrix}
    {\bf \Lambda}_{1} & \\
     & {\bf \Lambda}_{2}
  \end{matrix}\right],\,\,
  {\bf V} = [
  \begin{matrix}
    {\bf V}_{1} & {\bf V}_{2}
  \end{matrix}],
\end{equation}
where ${\bf \Lambda}_{1}=diag(\lambda_{1},\cdots,\lambda_{r})$, ${\bf V}_{1}=[\begin{matrix}
{\bf v}_{11} & \cdots & {\bf v}_{1r}  
\end{matrix}]$, and ${\bf \Lambda}_{2},\,\,{\bf V}_{2}$ are defined analogously.
We then define the rotated coordinates ${\bf z}_{1}\in\mathbb{R}^r$ and ${\bf z}_{2}\in\mathbb{R}^{d-r}$ by
\begin{equation}
  {\bf z}_1={\bf V}_{1}^{\mathrm{T}}{\bf x},\,\,{\bf z}_2={\bf V}_{2}^{\mathrm{T}}{\bf x}.
\end{equation}
One then takes ${\bf V}^\mathrm{T}_1$ as the low dimensional projection matrix and $\-{\bf z}_1$ as the dimension reduced variable. 
That is to say, one can now approximate the original function $f({\bf x})$ with a function, say $G(\-z_1)$, defined on 
the reduced space $\{\-z_1={\bf V}^{\mathrm{T}}_1 {\bf x}| {\bf x}\in\Omega\}$.
In particular we can define  function ${G}(\-z_1)$ as 
\begin{equation}
  {G}({\bf z}_1)=\mathbb{E}[f|{\bf z}_1] = \int_{{\bf z}_2}f({\bf V}_{1}{\bf z}_1+{\bf V}_{2}{\bf z}_2)\pi_{Z_{2}|Z_{1}}({\bf z}_2)d{\bf z}_2,
\end{equation}
and $G$ is the best mean-squared approximation of $f$ given ${\bf z}_{1}$ which follows from the so-called \emph{law of the unconscious statistician}~\cite{constantine2014active}. Thus an approximation of $f(\-x)$ can be constructed via ${G}(\-z_1)$:
\begin{equation}\label{as9}
  f({\bf x})\approx F({\bf x})\equiv{G}({\bf V}_{1}^{\mathrm{T}}{\bf x}).
\end{equation}
The following theorem provides an error bound for $F$ in terms of eigenvalues of ${\bf C}$~\cite{constantine2014active}:
\begin{theorem}\label{thm2.1}
  The mean squared error of F defined in (\ref{as9}) satisfies
  \begin{equation}
    \mathbb{E}[(f-F)^{2}]\leq  \alpha(\lambda_{r+1}+\cdots+\lambda_{d}),
  \end{equation}
  where $\alpha$ is a constant that depends on the domain $\mathcal{X}$ and the distribution $\pi$.
\end{theorem}

It should be clear that in practice it is usually not possible to evaluate Eq.(~\ref{asm1})  directly. Instead, one often approximates the integral via Monte Carlo
simulation. That is, assuming that the observed inputs are drawn from $\pi({\bf x})$, one approximates ${\bf C}$ using the observed gradients by:
\begin{equation}
  \widetilde{{\bf C}}=\frac{1}{N}\sum_{i=1}^{N}(\nabla_{{\bf x}}f({\bf x}^{(i)}))(\nabla_{{\bf x}}f({\bf x}^{(i)}))^{\mathrm{T}}.
\end{equation}
Next the eigenvalues and eigenvectors of $\widetilde{{\bf C}}$ are computed using the singular value decomposition (SVD). 
As is suggested in  \cite{constantine2014active}, the reduced dimensionality $r$
can be  determined by considering the spectrum of ${\bf C}_{N}$,
and one possible approach is to require that 
\begin{equation}
 \sum_{i=1}^r \lambda_i \geq \rho\sum_{i=1}^d \lambda_d, \label{e:rho}
 \end{equation}
for a prescribed ratio $\rho\in[0,1]$.
 We refer the readers to \cite{constantine2014active} for more details and other possible methods for determining $r$.
Next we can define,
\begin{equation}\label{mce:1}
   \hat{G}({\bf z}_1)=\frac{1}{N}\sum_{i=1}^{N}f({\bf V}_{1}{\bf z}_1+{\bf V}_{2}{\bf z}_2^{(i)}),
\end{equation}
where the ${\bf z}_2^{(i)}$'s are drown independently from the conditional density $\pi{(\-z_2|\-z_1)}$, and it follows that we 
can obtain an approximation of $f$,
\begin{equation}\label{as10}
  f({\bf x})\approx \hat{F}({\bf x})\equiv \hat{G}({\bf V}_{1}^{\mathrm{T}}{\bf x}).
\end{equation}
Similarly we can then derive an error bound for the Monte Carlo approximation $\hat{F}$, of $f$ in terms of eigenvalues of ${\bf C}$ as follows \cite{constantine2014active}:
\begin{theorem}\label{thm2.2}
  The mean squared error of $\hat{F}$ defined in Eq.~\eqref{as10} satisfies
  \begin{equation}
    \mathbb{E}[(f-\hat{F})^{2}]\leq  \alpha (1+\frac{1}{N})(\lambda_{r+1}+\cdots+\lambda_{d}),
  \end{equation}
  where $\alpha$ is a constant that depends on the domain $\mathcal{X}$ and the distribution $\pi$.
\end{theorem}
The proofs of Theorems~\ref{thm2.1} and \ref{thm2.2} can be found in \cite{constantine2014active} and will not be repeated here. 

\section{The Clustered Active subspace based local GP emulator}\label{sec:method}

\subsection{Clustered Active Subspace method} \label{sec:cas}

Using the AS method we can now construct a low dimensional subspace for function $f({\bf x})$. 
However, as has been discussed in Section~\ref{sec:intro}, in many practical problems with complex physics, a global low dimensional structure, i.e. 
one  that is applicable in the entire parameter domain, may not exist. 
Instead, different parameter regimes may have distinct low dimensional structures. 
 As a result, a global dimension reduction strategy may not apply. 
 In what follows we present a clustered AS method to conduct dimension reduction for such problems.

The main idea of our method is the following. 
Suppose that the function of interest admits different low dimensional structures in different  disjoint subdomains of the original input space, 
and ideally if we can identify these subdomains and the low dimensional subspace in each of them, 
  we can expect that the GP emulator constructed locally (i.e., to only use data points in the same subdomain as the new point of interest) enjoys a better performance than the emulator constructed globally.
First we can write the distribution $\pi(\-{\bf x})$  as a mixture:
\begin{equation}
\pi(\-{\bf x})= \sum_{j=1}^J w_j \pi_j(\-{\bf x}),\quad \sum_{j=1}^J{w_j}=1, \label{e:pimix}
\end{equation}
where each $\pi_j$ has a support $\Omega_j$ satisfying
\[\Omega_j\cap\Omega_{j'}=\emptyset \,\,\mathrm{for}\,\, \forall \,\,j\neq j',\qquad \cup_{j=1}^J \Omega_j=\Omega.\]
In what follows we refer each $\pi_j$ as a distribution cluster. 
Now we  assume that the function $f(\-{\bf x})$ has different properties in different region $\Omega_j$ and for each $\pi_j$ we can define
\begin{equation}
  {\bf C}_j=\mathbb{E}_{\pi_j({\bf x})}[(\nabla_{{\bf x}}f)(\nabla_{{\bf x}}f)^{\mathrm{T}}]
\end{equation}
and compute the associated ``clustered'' active subspace accordingly. 
By going through the same procedure as described in Section~\ref{sec:as}, we can obtain a low dimensional projection matrix 
$\-V_{1,j}$ and its compliment $\-V_{2,j}$
for each distribution cluster $\pi_j$.  
As a result we obtain a set of $J$ low-dimensional projections $\{\-V_{1,j}\}_{j=1}^J$,
and if the function's structure is very different with respect to different clusters/subdomains,
these low-dimension projections are significantly different from each other. 
Defining  $\-z_{1,j}=\-V^\mathrm{T}_{1,j}\-x$ and  $\-z_{2,j}=\-V^\mathrm{T}_{2,j}\-x$ for $i=1,\cdots,J$,
 we obtain a set of local approximations, 
\begin{equation}
  {G}_j({\bf z}_{1,j})=\mathbb{E}_{\pi_j}[f|{\bf z}_{1,j}] = \int_{{\bf z}_{2,j}}f({\bf V}_{1,j}{\bf z}_{1,j}+{\bf V}_{2,j}{\bf z}_{2,j})
  \pi_j{(\-z_{2,j}|\-z_{1,j})}d{\bf z}_{2,j},
\end{equation}
and for any $\-x\in\Omega_j$,
\begin{equation}
  f({\bf x})\approx {F}_j({\bf x})\equiv {G}_j({\bf V}_{1,j}^{\mathrm{T}}{\bf x}).
\end{equation}
Consequently we can define a global approximation for $f(\-x)$ as,
\begin{equation}
  f({\bf x})\approx F({\bf x})=\sum_{j=1}^{J}{F}_{j}({\bf x})I_{\Omega_j}(\-x), \label{e:globalF}
\end{equation}
where $I_{\Omega_j}$ is an indicator function,
\begin{equation}\label{e:ind}
	I_{\Omega_j}({\bf x})=\left\{
	\begin{aligned}
		1,&\,\,{\bf x}\in\Omega_{j},\\
		0,&\,\, {\bf x}\notin\Omega_{j}.
	\end{aligned}\right.
\end{equation}
The error bound of $F$ defined in Eq.~\eqref{e:globalF} is given by the following theorem. 
\begin{theorem}\label{thm3.1}
Let $\{\lambda_{i,j}\}_{i=1}^d$ be the eigenvalues of $C_j$ in a descending order
and $r_j$ be the reduced dimensionality associated with the $j$-th cluster. The mean squared error of $F$ defined in Eq.~\eqref{e:globalF} satisfies
  \begin{equation}
    \mathbb{E}_{\pi}[(f-F)^{2}]\leq  \sum_{j=1}^{J}\alpha_{j}(\lambda_{r_{j}+1,j}+\cdots+\lambda_{d,j}),
  \end{equation}
  where  $\alpha_{j}$'s are constants that depend on the domains  $\Omega_j$'s and the distribution clusters $\pi_{j}$'s.
\end{theorem}

Next we shall establish the error analysis for the Monte Carlo estimation as is done for the AS method. 
First suppose that we have samples $\-x^{(1)},\cdots,x^{(N)}$ drawn from the distribution cluster $\pi_j$, 
and we can estimate the sample variance,
\begin{equation}
  \widetilde{{\bf C}}_j=\frac{1}{N}\sum_{i=1}^{N}(\nabla_{{\bf x}}f({\bf x}^{(i)}))(\nabla_{{\bf x}}f({\bf x}^{(i)}))^{\mathrm{T}},
\end{equation}
and once again we can obtain the two matrices $\-V_{1,j}$ and $\-V_{2,j}$ associated with $\widetilde{\bf C}_j$. 
Let  $\-z_{1,j}=\-V^\mathrm{T}_{1,j}\-x$ and  $\-z_{2,j}=\-V^\mathrm{T}_{2,j}\-x$ and we get
\begin{equation}
	 \hat{G}_{k}({\bf z}_1^{(k)})=\frac{1}{N_{j}}\sum_{i=1}^{N_{j}}f({\bf V}_{1}^{k}{\bf z}_1^{(k)}+{\bf V}_{2}^{k}{{\bf z}_2^{(k)}}^{i}),
\end{equation}
in which ${{\bf z}_2^{(k)}}^{i}$ are drawn i.i.d. from the conditional distribution $\pi_j{(\-z_{2,j}|\-z_{1,j})}$.
Finally we obtain an approximation $\hat{F}$ of $f$ are as follows:
\begin{equation}\label{mce:2}
	f({\bf x})\approx \hat{F}({\bf x})=\sum_{j=1}^{J}\hat{F}_{j}({\bf x})I_{\Omega_j}(\-x),
\end{equation}
where
\begin{equation}
	\hat{F}_{j}({\bf x}) =  \hat{G}_j({\bf V}_{1,j}^{\mathrm{T}}{\bf x}),
	\end{equation}
	and $I_{\Omega_j}(\cdot)$ is the indicator function defined in Eq.~\eqref{e:ind}.
Then we can derive an error bound for the Monte Carlo approximation $\hat{F}$ of $f$ which is stated by the following theorem:
\begin{theorem}\label{thm3.2}
  The mean squared error of $\hat{F}$ defined in Eq.~\eqref{mce:2} satisfies
  \begin{equation}
    \mathbb{E}[(f-\hat{F})^{2}]\leq  \sum_{j=1}^{J}\alpha_{j}(1+\frac{1}{N_{j}})(\lambda_{r_{j}+1}+\cdots+\lambda_{d}),
  \end{equation}
  where $r_j$ denotes the reduced dimensionality in $\Omega_j$ and $\alpha_{j}$'s are constants that depend on the domains $\Omega_j$'s and the distributions $\pi_{j}$'s.
\end{theorem}
In the Appendix  we provide a proof for Theorem~\ref{thm3.2} and the proof for Theorem~\ref{thm3.1} proceeds similarly and thus is omitted.

\subsection{Gradient-based data clustering} \label{sec:clustering}
In Section~\ref{sec:cas}, we have provided the main framework of the clustered active subspace method. 
Clearly a key issue yet to be addressed is that the distribution clusters are not known in advance, and 
we need to identify the distribution clusters that represent different function structures. 
In particular in Section~\ref{sec:cas} we have assumed that it is known in advance that
the samples ${\bf x}^{(1)},\cdots,{\bf x}^{(N_j)}$  are drawn from each distribution cluster $\pi_j(\bf x)$.
However, in reality all the samples are drawn from the distribution $\pi$ and we need to 
partition them into different data clusters, which implicitly defines the distribution clusters as well. 
In other word, we do not need to explicitly obtain the distribution clusters, and for implementation purpose 
we only need to cluster the data points.

The main idea here is to cluster the data points based on the gradient information which may reveal the local structure of the function $f(\-x)$. 
For this purpose we choose to use the hierarchical clustering algorithm, which groups data over a variety of scales by creating a cluster tree or dendrogram. 
An important feature of hierarchical clustering is that the tree is not a single set of clusters, but rather a multilevel hierarchy, where clusters at one level are joined as clusters at the next level. This feature is important as it allows users to decide the level or scale of clustering based on the specific application.
We refer to~\cite{kaufman2008agglomerative} for more details of the hierarchical clustering methods.

In  hierarchical clustering, a key is to define an appropriate distance that can measure the similarity or dissimilarity  between the data points,
and this distance should represent the properties of the data points that users hope to distinguish. 
In our problem, as has been mentioned, we expect that different clusters represent different low-dimensional structures of the function, which
in the AS framework is encoded in the gradient information. 
This motivates us to cluster the data based on the gradient. 
To do so, we  define the absolute cosine distance measure as follows:
\begin{equation}
  \Delta({\bf x},{\bf x}') = 1-|\cos({\bf g_{x}},{\bf g_{x'}})|, \label{e:cos}
\end{equation}
where ${\bf g}_{{\bf x}},{\bf g}_{{\bf x}'}$ are the corresponding gradient of ${\bf x},{\bf x}'$,
and we conduct a clustering of the data points based on this distance.
This distance function can be further generalized, taking into account of the Euclidean distance between points and combining it with the gradient similarity.  
Namely we set the distance to be
\begin{equation}
  \Delta_\eta({\bf x},{\bf x}') = \eta*(1-|\cos({\bf g_{x}},{\bf g_{x}}')|)+(1-\eta)*(||{\bf x}-{\bf x}'||_{2}/\sqrt{d}),\label{e:pdist}
\end{equation}
where $\eta\in[0,1]$ is a hyper-parameter to balance the influence between the discrepancy on locations and gradients of ${\bf x}$ and $d$ is the dimension of ${\bf x}$. We divide the euclidean distance by $\sqrt{d}$ to adjust the second term to be in the same level of scales with the first absolute cosine distance.
It should also be clear  that, when $\eta=1$, the distance is reduced to the absolute cosine distance in Eq.~\eqref{e:cos} which only uses the gradient information. 
Indeed, in most of the problems, we can simply use the gradient-based distance measure, and 
      the  Euclidean distance is introduced  as an insurance policy to prevent certain extreme scenarios where the gradient based distance may fail. 
      Thus in practice we recommend to choose $\eta$ to be close to 1.

Another important issue in hierarchical clustering is to choose the linkage criterion, and in our numerical tests we have found that the
unweighted average linkage clustering ~\cite{kaufman2008agglomerative} has the best performance overall thus is used in this work. That said, the method proposed does not reply on any specific choice of the linkage function. 
Next we discuss the procedure of the gradient based clustering. 
We modify the notation defined by Eq.~\eqref{e:tset}, extending the training set to include the gradient:
\begin{equation}
D=\{({\bf x}^{(n)},\,y^{(n)},\,{\bf g}^{(n)})|y^{(n)}=f({\bf x}^{(n)}),\,\,{\bf g}^{(n)}=\nabla_{{\bf x}} f({\bf x}^{(n)})\}_{n=1}^N.\label{e:tset2}
\end{equation}
and present the complete procedure of the clustered active subspace method in Algorithm~\ref{alg:cas}.
Finally  one can see that  the CAS method requires the number of clusters $J$ as an input, and in our method it is determined 
via a $k$-fold cross-validation~\cite{arlot2010survey} where details are provided in \ref{sec:cv}. 

\begin{algorithm}[ht]
    \caption{Clustered active subspace dimensionality reduction(CAS)}\label{alg:cas}
    \begin{algorithmic}
    \Require{ Training set $D$; the number of clusters $J$;  the dimension reduction ratio $\rho$. }
    \Ensure{  $\{D_j,\,{\bf V}_{1,j}\}_{j=1}^{J}$, where $D_j$ are the data clusters and ${\bf B}_{j}$ are the associated DR projection matrices.}
    \State $ \{D_j\}_{j=1}^J\leftarrow$ cluster data set $D$ into $J$ clusters according to distance~\eqref{e:pdist};
    \For{$j=1,\cdots,J$}
    
    \State  ${\bf C}_j = \frac{1}{card(D_{j})}\sum_{({\bf x}^{(n)},y^{(n)},{\bf g}^{(n)})\in D_{j}}({\bf g}^{(n)})({\bf g}^{(n)})^{\mathrm{T}}$;
    \State Conduct SVD to matrix ${\bf C}_j$  obtaining ${\bf C}_j = {\bf V}_{j}{\bf \Lambda}_{j} {\bf V}_{j}^{\mathrm{T}}$;
    \State Determine the reduced dimensionality $r_j$  based on ratio $\rho$;
   \State $ {\bf V}_{1,j}=[\begin{matrix}
{\bf v}_{j1} & \cdots & {\bf v}_{jr_j}   
\end{matrix}]$;\;
    \EndFor
    \State return $\{D_j, {\bf V}_{1,j}\}_{j=1}^{J}$\;
    \end{algorithmic}
\end{algorithm}


\subsection{Local GP emulator with clustered dimensionality reduction}
In Section~\ref{sec:clustering} we have discussed the method to obtain the data clusters as well as the dimension reduction projection matrices associated to them. 
To make use of them, another key issue is  to determine which cluster a new point belongs to,
which can be posed as a classification problem. 
It is important to note that, practically we only have the knowledge of the input parameter itself, and we do not know
the function value or the gradient. 
Since the training data points have been clustered and labelled, we construct a classifier based on the parameter value $\-{\bf x}$ only, in a supervised manner. 
In particular we choose to construct the classifier with the Support Vector Machine (SVM) model~\cite{vapnik2013nature}
mainly for its simplicity, while noting that other classification methods can also be used here. 
To start, we construct the data set for the classification problem as 
\begin{equation}
D_c=\{({\bf x}^{(n)}, b^{(n)})\}_{n=1}^N,
\end{equation}
where $b^{(n)}$ is the label assigned to cluster $D_j$ if ${\bf x}^{(n)}\in D_{j}$.
Using the classification data set $D_c$ we are able to train a SVM classifier denoted as $b=\mathrm{SVM}({\bf x})$.
We omit the training procedure of the SVM model and interested readers may consult ~\cite{vapnik2013nature}. 
Now for any given new point ${\bf x}^*$, we first use the trained SVM model to determine which cluster it belongs to,
and then we conduct dimension reduction of all the data points in the chosen cluster with the associated dimension reduction projection matrix. 
Finally a local GP emulator is constructed  with the dimension-reduced data points for the new point $\-x^*$. 
More precisely, suppose that the cluster predicted by the SVM model is $D_{j^*}$, with dimension reduction matrix ${\bf V}_{1,j^*}$,
and we define the following data set: 
\begin{equation}
D_{GP} = \{({\bf z}^{(i)},y^{(i)})| {\bf z}^{(i)} = {\bf V}_{1,j^*}^\mathrm{T} {\bf x}^{(i)},\,\,\forall {\bf x}^{(i)} \in D_{j^*}\},
\end{equation}
  for the construction of the GP model. 
Finally  the GP model is constructed using $D_{GP}$ with the procedure outlined in Section~\ref{sec:gp},  
which is then used to predict the value of $f({\bf x}^*)$.
We reinstate here that the GP model is constructed on the low dimensional subspace obtained by ${\bf V}_{1,j^*}$ and only with data points in $D_{j^*}$,
and so it is regarded as a local GP model. 
We summarize the complete procedure for constructing such a local GP model in Alg.~\ref{alg:lgp}.
Finally a number of remarks are listed in order:

\begin{itemize}
\item An important question is the number of data points used in the method. While noting that the actual number of data points needed is problem-dependent, we reinstate that 
here we mainly consider the situation where the number of data points is typically limited, i.e., insufficient to conduct regression 
in the high-dimensional space. The dimension reduction technical should be able to improve the regression performance in
such a  situation. 
\item Compared to the standard AS, 
the proposed CAS method is more computationally expensive mainly because the additional clustering procedure is needed. 
However, as has been mentioned earlier,  we consider in this work problems with exceedingly expensive computer models, and as such 
the dimension reduction procedure is not a main contributor to the total computational cost, as long as it does not involve simulating the computer model.
\end{itemize}

\begin{algorithm}[H]
\caption{The construction of the low dimensional local GP model}\label{alg:lgp}
\begin{algorithmic}
	\Require{ $D_{c}$ classification data set; test point $x^{*}$. }
    \Ensure{ testing prediction $y^{*}$. }
    \State Train a SVM model based on the training set $D_c$, denoted as $b=\mathrm{SVM}({\bf x})$;\;
    \State    Compute $b^*=\mathrm{SVM}({\bf x}^*)$ and let $D_{j^*},{\bf V}_{1,j^*}$ respectively be the data cluster and projection matrix corresponding to label $b^*$;\;
    \State Obtain the training set for GP: 
    $$D_{GP} = \{({\bf z}^{(i)},y^{(i)})| {\bf z}^{(i)} = {\bf V}_{1,j^*}^\mathrm{T} {\bf x}^{(i)},\,\,\forall {\bf x}^{(i)} \in D_{j^*}\};$$
    \State Construct the GP model $f_{j^*}$ with data set $D_{GP}$, and use $f_{j^*}({\bf V}_{1,j^*}^\mathrm{T}{\bf x}^*)$ to predict the value of $f({\bf x}^*)$.
\end{algorithmic}  
\end{algorithm}


\section{Numerical results}\label{sec:numeres}

\subsection{Example 1: a piece-wise function}

\begin{figure}
    \begin{minipage}[t]{0.5\linewidth}
    \small
    \centering
    \includegraphics[width=1\linewidth]{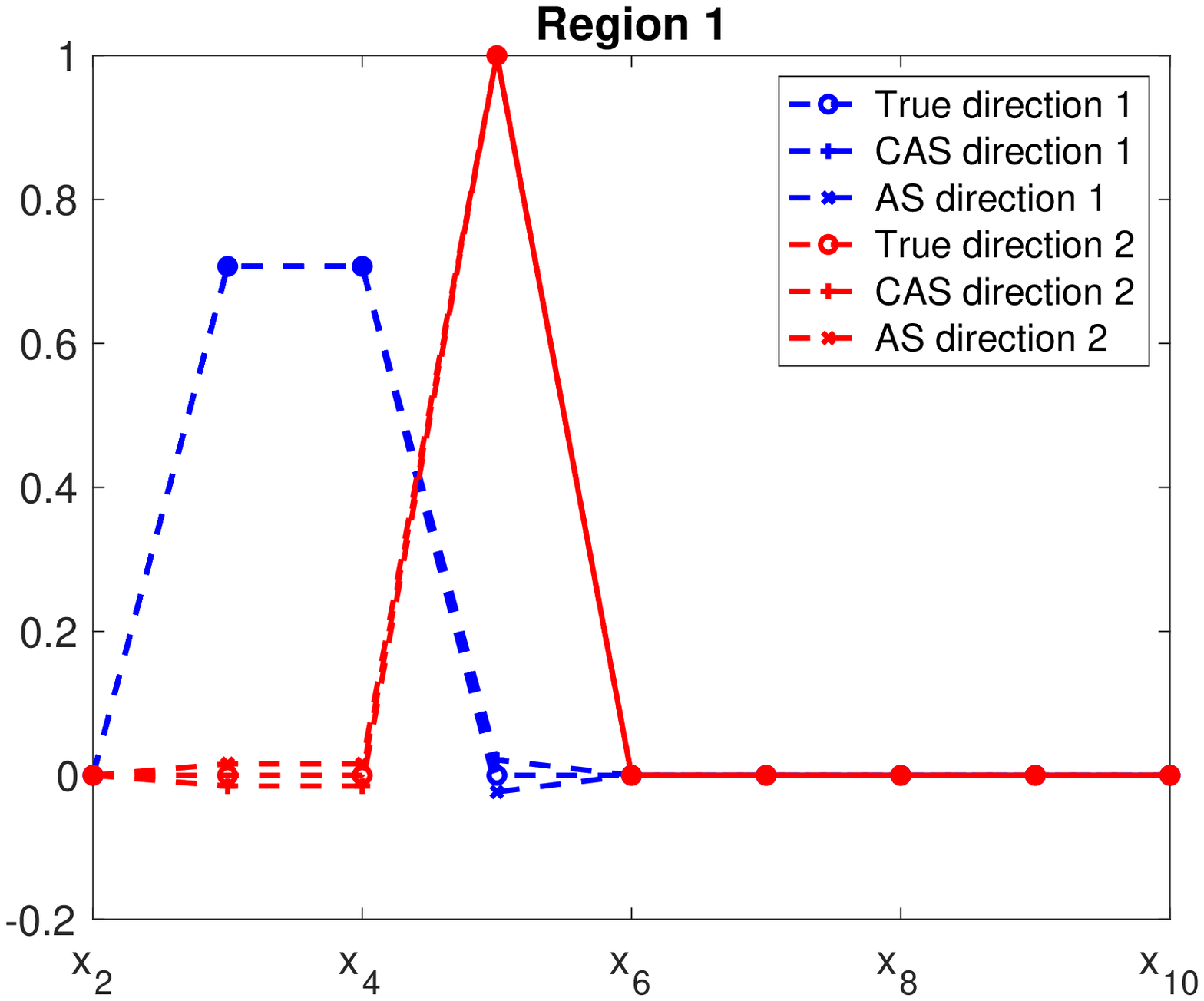}
    \end{minipage}
    \hfill
    \begin{minipage}[t]{0.5\linewidth}
    \small
    \centering
    \includegraphics[width=1\linewidth]{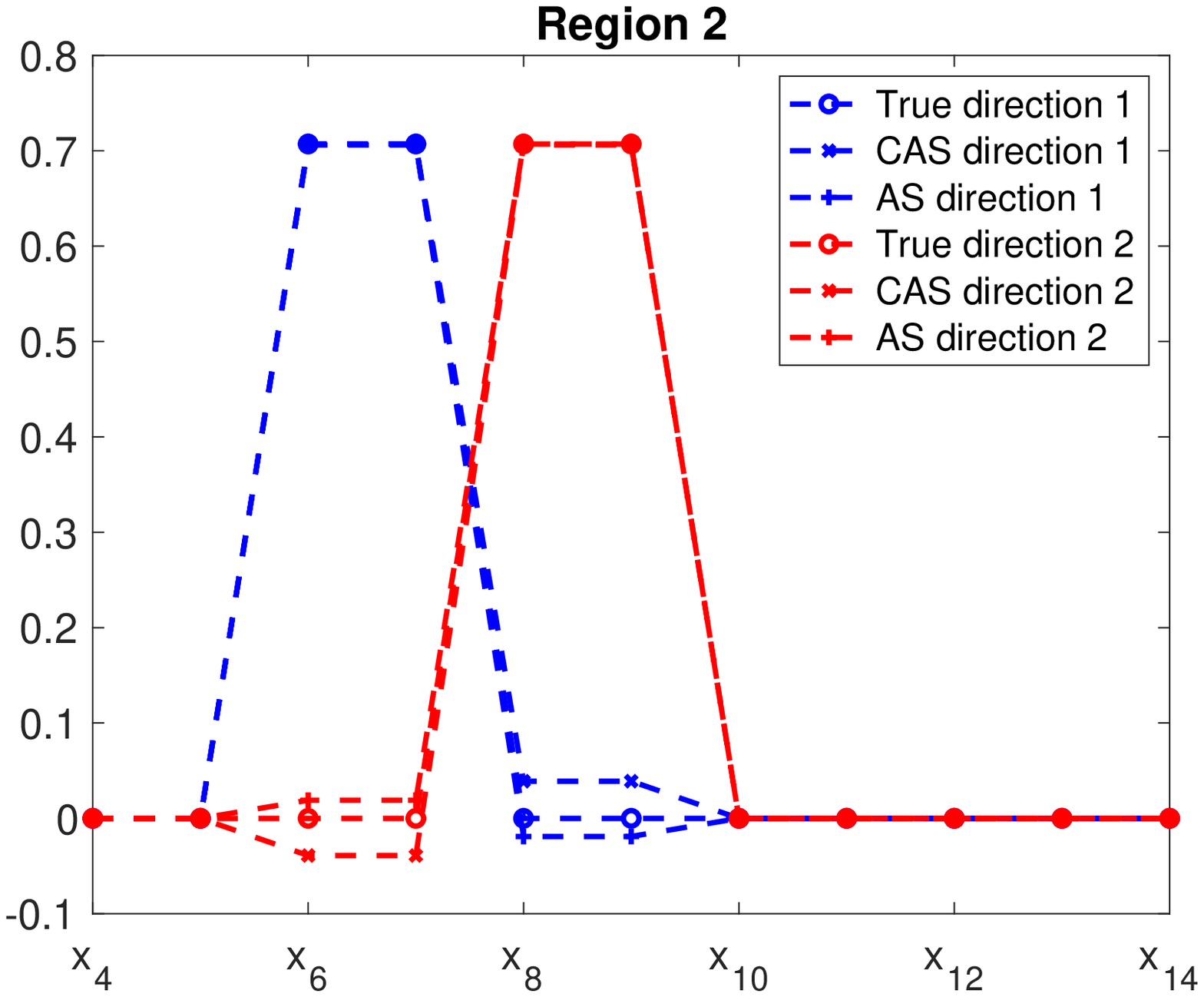}
    \end{minipage}
    \vfill
    \begin{minipage}[t]{0.5\linewidth}
    \small
    \centering
    \includegraphics[width=1\linewidth]{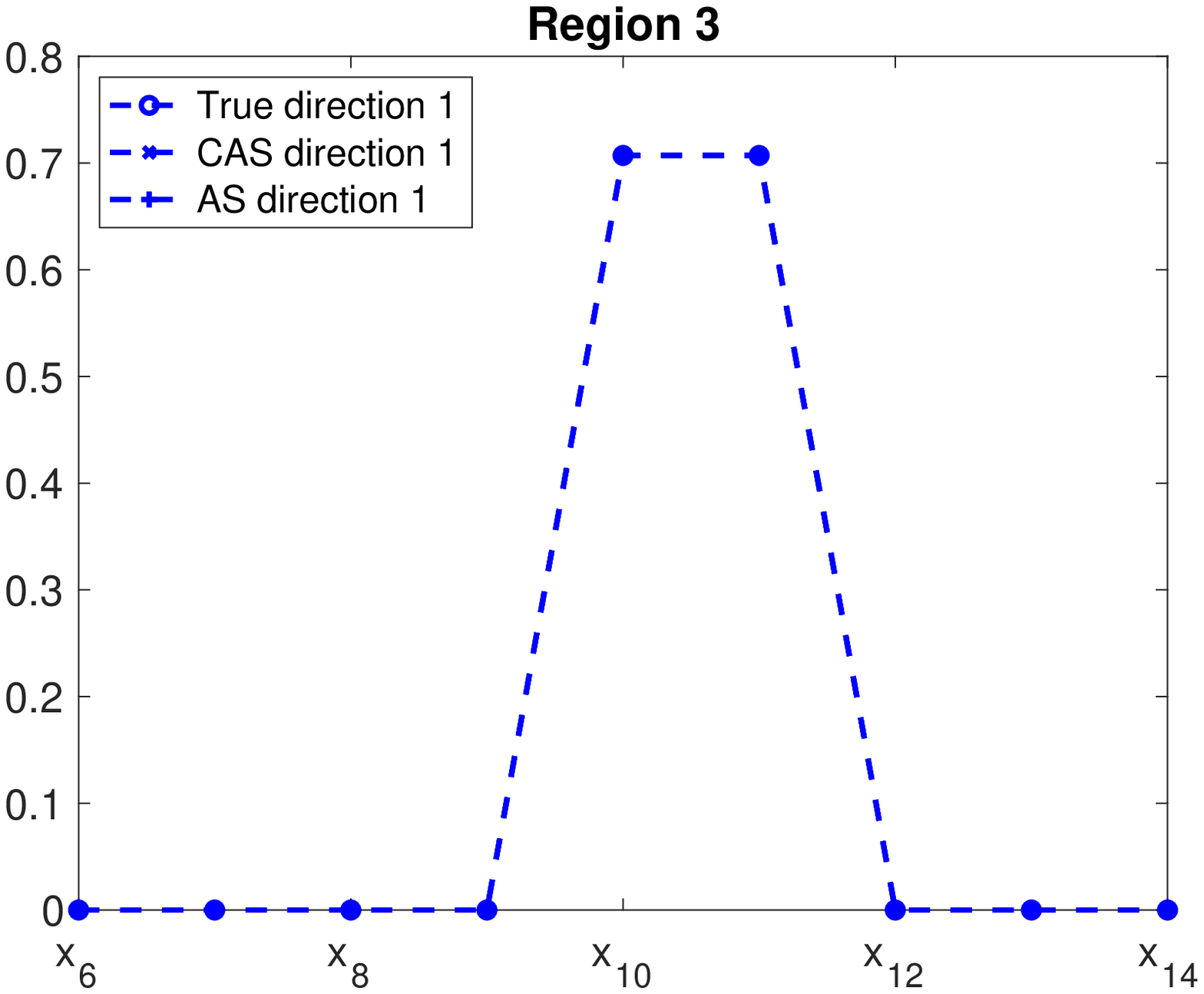}
    \end{minipage}
    \hfill
    \begin{minipage}[t]{0.5\linewidth}
    \small
    \centering
    \includegraphics[width=1\linewidth]{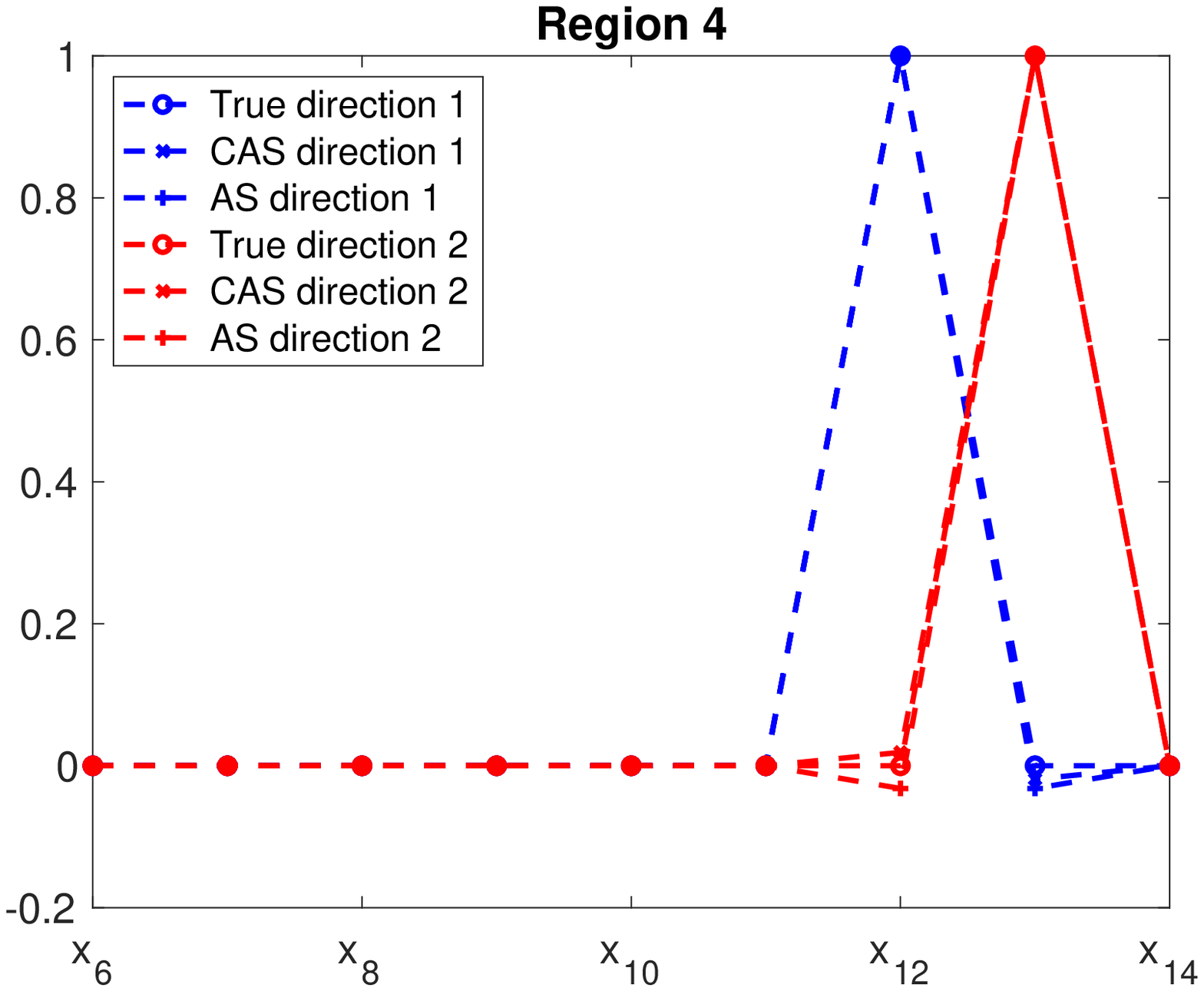}
    \end{minipage}
    	\caption{DR directions in four regions.}\label{f:fun1dr}
\end{figure}

To illustrate the effectiveness of the proposed approach, we first consider a piece-wise function which has four distinct
low dimensional structures in four disjoint  regions. Specifically the function is defined on ${\bf x}\in[-1,1]^{50}$, and admits the form, 
\begin{equation}\label{e:fun1}
  f({\bf x})=\left\{
  \begin{aligned}
    & (1+x_{3}+x_{4})x_{5}, & x_{1}<0 \,\,and\,\, x_{2}<0~(\mathrm{Region 1}),\\
    & (1+x_{6}+x_{7})(x_{8}+x_{9}), & x_{1}<0 \,\,and\,\, x_{2}\geq0~(\mathrm{Region 2}),\\
    & 1+x_{10}+x_{11}, & x_{1}\geq0 \,\,and\,\, x_{2}<0~(\mathrm{Region 3}),\\
    & (1+x_{12})x_{13}, & x_{1}\geq0 \,\,and\,\, x_{2}\geq0~(\mathrm{Region 4}).\\
  \end{aligned}\right.
\end{equation}
This function relies on completely different dimensions of the input space in the four different regions, and
we want to test if the ability to identify these regions and their individual low dimensional structures 
 can improve the performance of the GP emulator.
 Also since the low dimensional structure of this problem is analytically available,
 we can use it to validate the dimension reduction results. 
To this end we  hope  that the proposed CAS method can correctly identify the regions (in the form of clusters) and the reduced dimensions in each of them. 
In the numerical tests,  the distribution of $\-x$ is taken to be uniform defined on ${\bf x}\in[-1,1]^{50}$, 
and 1000 training samples and 5000 testing samples are drawn from the distribution.
Our experiments involves two main steps: conducting a dimension reduction and  constructing the GP model in the low dimensional subspace. 
First we want to examine if the standard AS and the CAS methods  can correctly identify the DR directions of this function (one can see from
Eq.~\eqref{e:fun1} that the function admits totally 7 actual DR directions).
To do so, we compare the actual DR directions with those identified with both AS and CAS (with 4 clusters) methods:
for regions 1, 2 and 4,  each has two DR directions and region 3 only has one.
In Fig.~\ref{f:fun1dr} we show the DR directions in all the four regions,
and for the regions with two DR directions they are marked with different colors blue and red. 
As one can see from the figure, the directions computed with both AS and CAS agree well with the actual ones,
while, in addition to the DR directions, CAS can also correctly identify the clusters or regions where the low dimensional structures are
different. 

Next we shall demonstrate that the ability of CAS to identify these different clusters can help improve the performance of the GP emulator. 
In addition to the CAS-LGP algorithm, we also employ AS, SIR~\cite{li1991sliced}  and SAVE~\cite{doi:10.1080/01621459.1991.10475036} to reduce the dimensionality globally and then construct the GP emulator in the resulting dimension-reduced space.
For comparison purposes, we also conduct the test of direct GP emulator in the original space, without dimension reduction. 
In CAS-LGP, we consider two cases of the distance function used in the clustering: $\eta=1$ and $\eta=0.5$, where the first case 
only considers the gradient information while the 2nd combines gradient similarity with the Euclidean distance. 
In all the examples, the predictive performance of GP emulator is measured by the normalized mean-square-error (NMSE):
\begin{equation}
  NMSE = \frac{\sum_{n=1}^{N}(f({\bf x}^{(n)})-\hat{f}({\bf x}^{(n)}))^{2}}{\sum_{n=1}^{N}f({\bf x}^{(n)})^{2}},
\end{equation}
where ${\bf x}^{(1)},\cdots,{\bf x}^{(N)}$ are test samples drawn from the distribution $\pi({\bf x})$, 
$f({\bf x})$ is the actual function and $\hat{f}$ is the posterior mean of the GP model.
For the global dimension reduction methods, we  test five different numbers of reduced dimensions: $r=1,\,2,\,3,\,4,\,7$ (we choose 
7 because the total number of intrinsic dimensions is 7),
and for CAS-LGP we compute the results with 2 to 4 clusters each with $r=1,\,2,\,3,\,4$ reduced dimensions,
where we note that when the number of clusters is taken to be $1$, the method reduces to the standard AS.
We compute the NMSE for each case and  summarize all the results in Table~\ref{tb:nmse1}.
First of all, as we can see from the table, for all the cases where the reduced dimensionality 
is from 1 to 4, CAS-LGP with four clusters yield by far the best results. 
Moreover, interestingly the table also illustrates that, even though the standard AS can correctly identify the 7 DR directions, the resulting GP emulator is 
severely  inaccurate, which is even less accurate than the GP constructed in the original space. 
Lastly we note that the distance function in this example seems to have little impact on the performance, as 
the results of the two cases ($\eta=1$ and $\eta=0.5$) are nearly identical. 
This example demonstrates that, if the model of interest has clearly different low dimensional structure in different regions, 
the proposed method can effectively identify these regions and their associated low dimensional structures,
which in turn produces an accurate local GP emulator.

    \begin{table}
      \centering
      \caption{The NMSE results of the piece-wise example.}\label{tb:nmse1}
      \begin{tabular}{|c|c|c|c|c|c|c|c|}
      \hline
      \multicolumn{8}{|c|}{$\eta=1$}\\
      \hline
      Method & \multicolumn{3}{|c|}{CAS-LGP} & \multirow{3}*{AS-GP} & \multirow{3}*{SIR-GP} & \multirow{3}*{SAVE-GP} & \multirow{3}*{GP}\\
      \cline{1-4}     
      \diagbox{d}{k} & 2 & 3 & 4 & & & &\\
      \hline
      1 & 73.1\% & 68.5\% & {\bf 41.1\%} & 91.6\% & 72.6\% & 88.6\% & \multirow{5}*{67.5\%} \\
      \cline{1-7} 
      2 & 72.1\% & 57.6\% & {\bf 20.3\%} & 86.1\% & 73.2\% & 85.3\% &  \\
      \cline{1-7} 
      3 & 65.1\% & 56.6\% & {\bf 20.3\%} & 85.3\% & 72.7\% & 86.0\% &  \\
      \cline{1-7} 
      4 & 61.9\% & 55.7\% & {\bf 20.3\%} & 80.5\% & 72.2\% & 86.2\% &  \\
      \cline{1-7} 
      7 &  &  & & 81.7\%  & 71.6\% & 82.5\% &  \\
      \hline
      \multicolumn{8}{|c|}{$\eta=0.5$}\\
      \hline
      Method & \multicolumn{3}{|c|}{CAS-LGP} & \multirow{3}*{AS-GP} & \multirow{3}*{SIR-GP} & \multirow{3}*{SAVE-GP} & \multirow{3}*{GP}\\
      \cline{1-4}     
      \diagbox{d}{k} & 2 & 3 & 4 & & & &\\
      \hline
      1 & 73.1\% & 68.5\% & {\bf 41.1\%} & 91.6\% & 72.6\% & 88.6\% & \multirow{5}*{67.5\%} \\
      \cline{1-7} 
      2 & 72.1\% & 57.6\% & {\bf 20.3\%} & 86.1\% & 73.2\% & 85.3\% &  \\
      \cline{1-7} 
      3 & 65.1\% & 56.6\% & {\bf 20.3\%} & 85.3\% & 72.7\% & 86.0\% &  \\
      \cline{1-7} 
      4 & 61.9\% & 55.7\% & {\bf 20.3\%} & 80.5\% & 72.2\% & 86.2\% &  \\
      \cline{1-7} 
      7 &  &  & & 81.7\%  & 71.6\% & 82.5\% &  \\
      \hline
      \end{tabular}
      
    \end{table}

\subsection{A Gaussian mixture}
\begin{figure}
	\caption{Principal directions in three different regimes}\label{f:dr4gm}
    \begin{minipage}[t]{0.5\linewidth}
    \small
    \centering
    \includegraphics[width=1\linewidth]{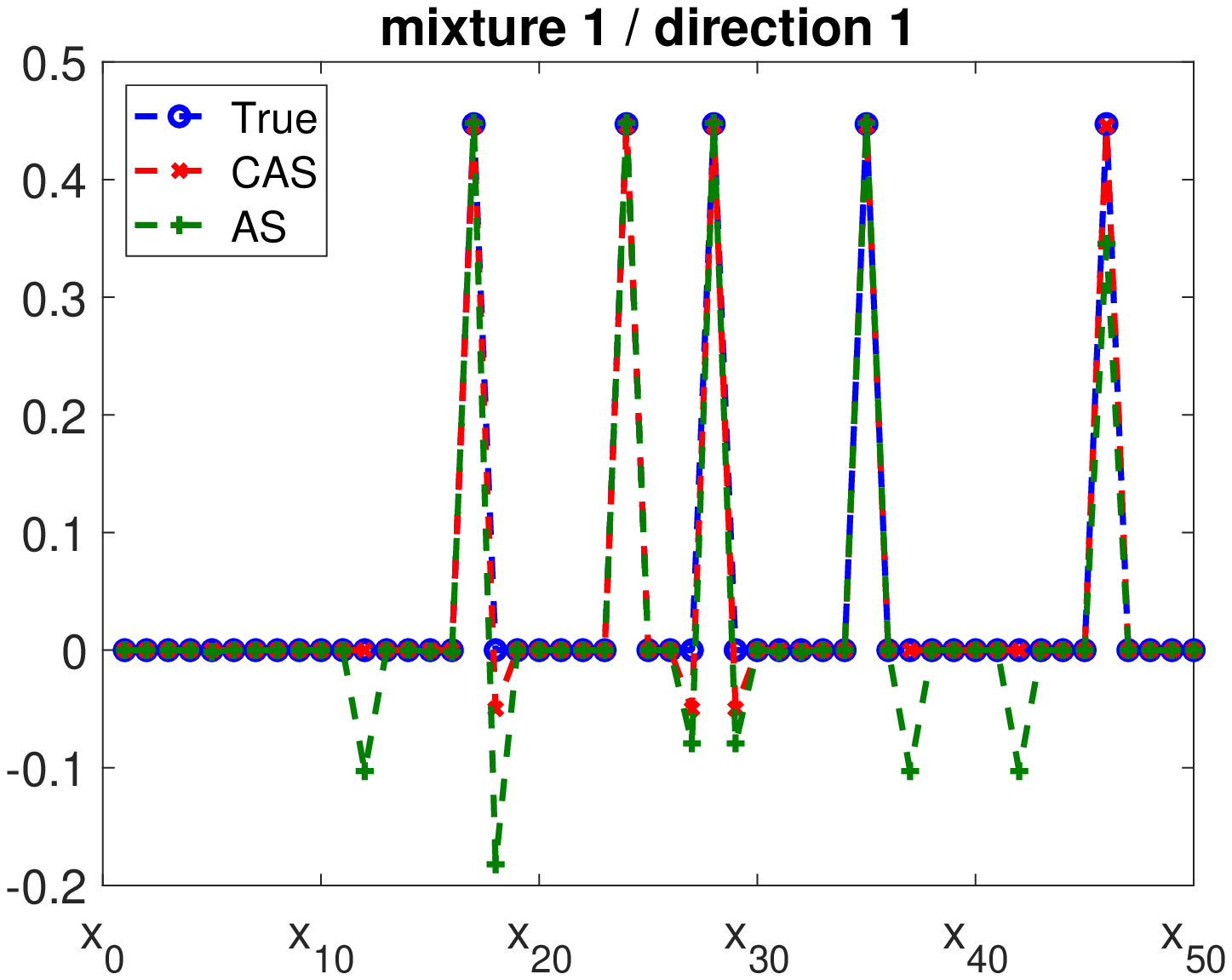}
    \end{minipage}
    \hfill
    \begin{minipage}[t]{0.5\linewidth}
    \small
    \centering
    \includegraphics[width=1\linewidth]{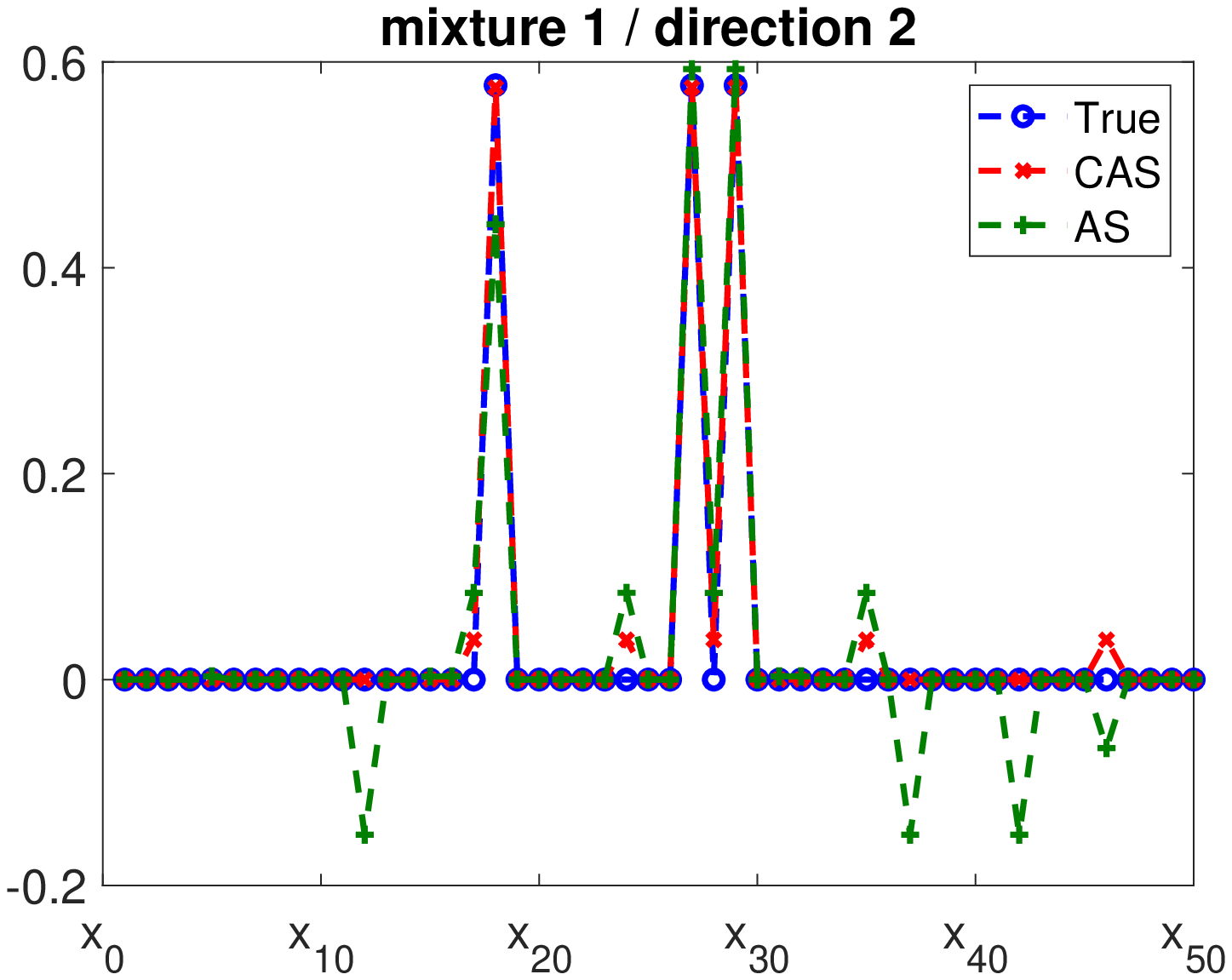}
    \end{minipage}
    \vfill
    \begin{minipage}[t]{0.5\linewidth}
    \small
    \centering
    \includegraphics[width=1\linewidth]{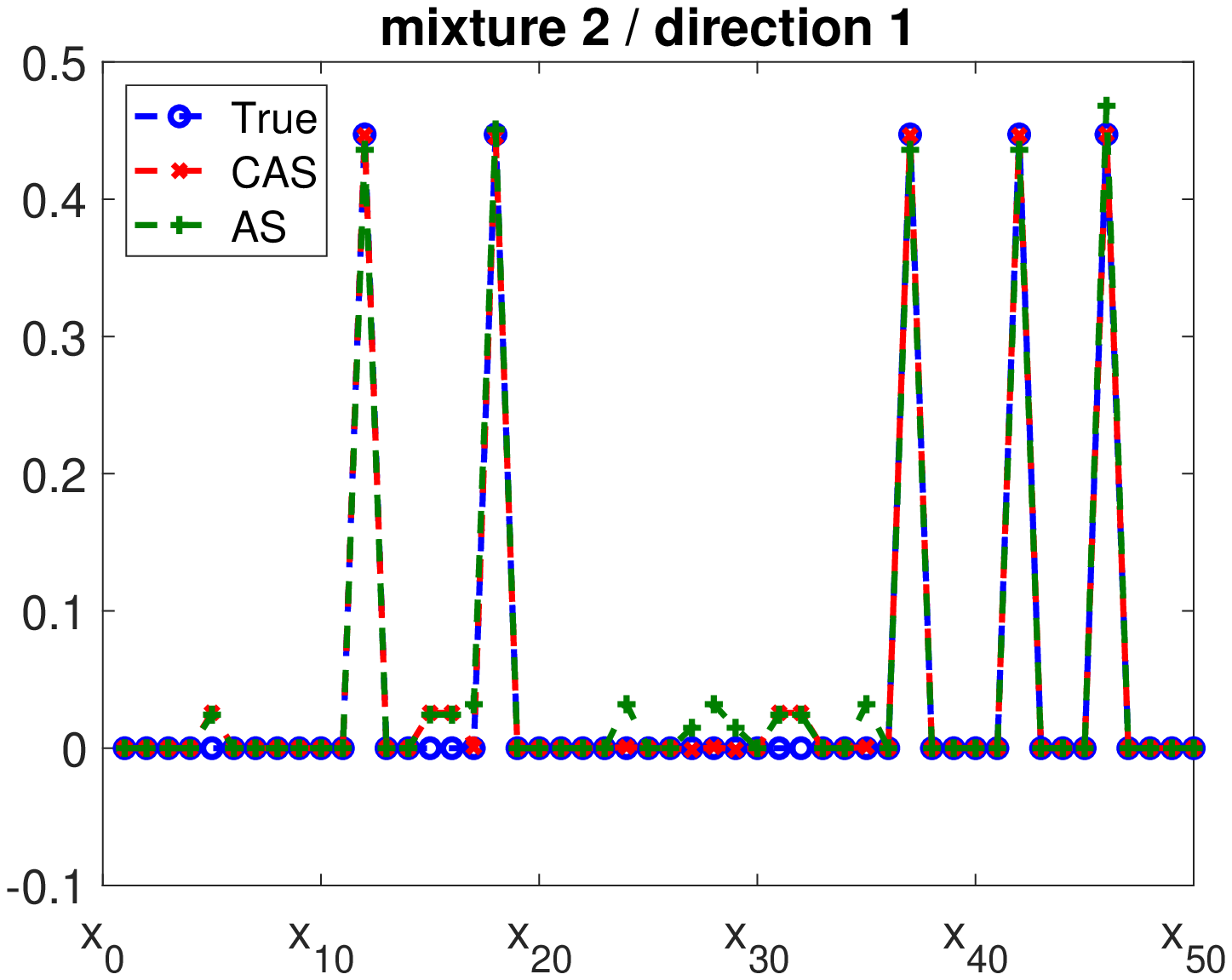}
    \end{minipage}
    \hfill
    \begin{minipage}[t]{0.5\linewidth}
    \small
    \centering
    \includegraphics[width=1\linewidth]{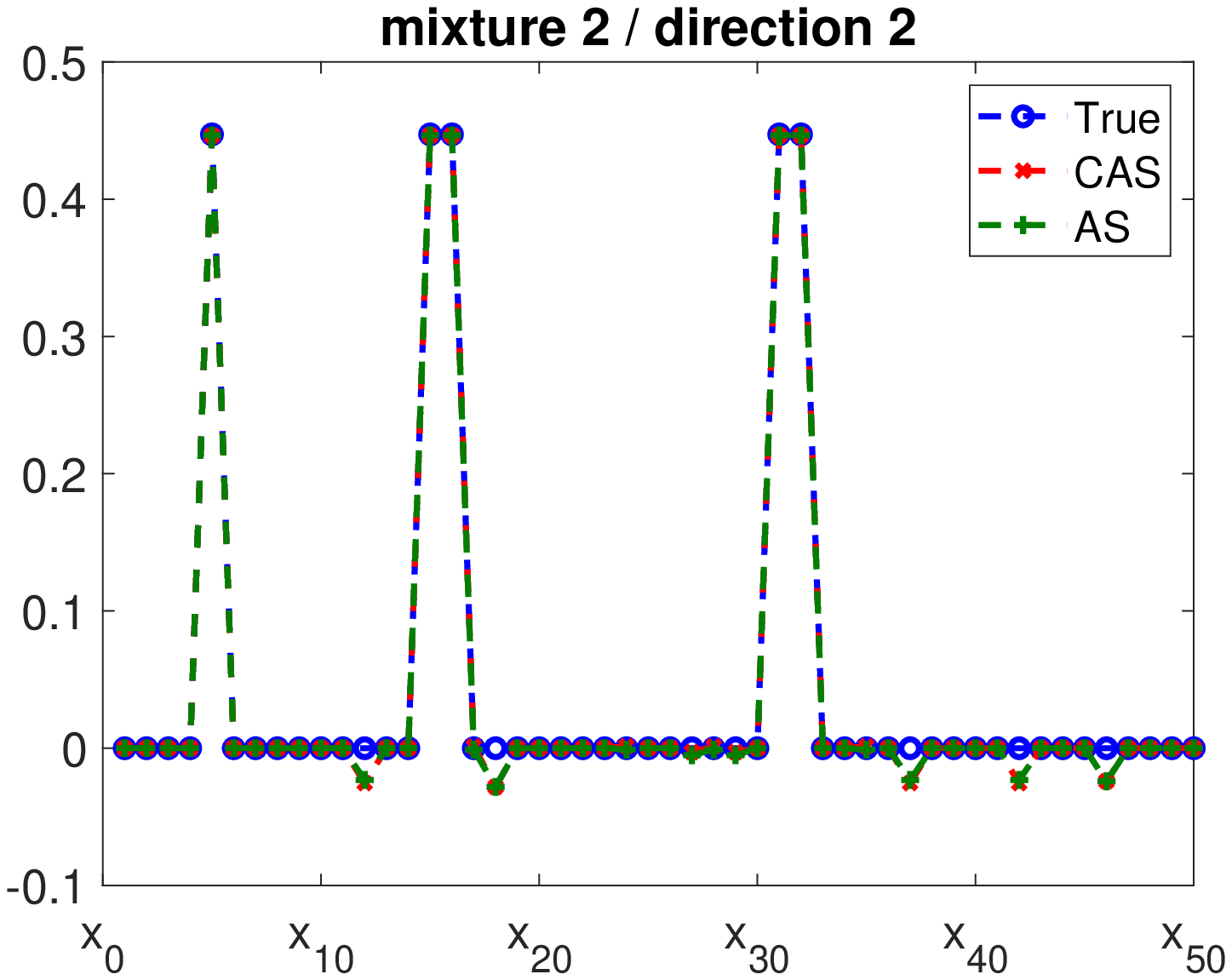}
    \end{minipage}
     \begin{minipage}[t]{0.5\linewidth}
    \small
    \centering
    \includegraphics[width=1\linewidth]{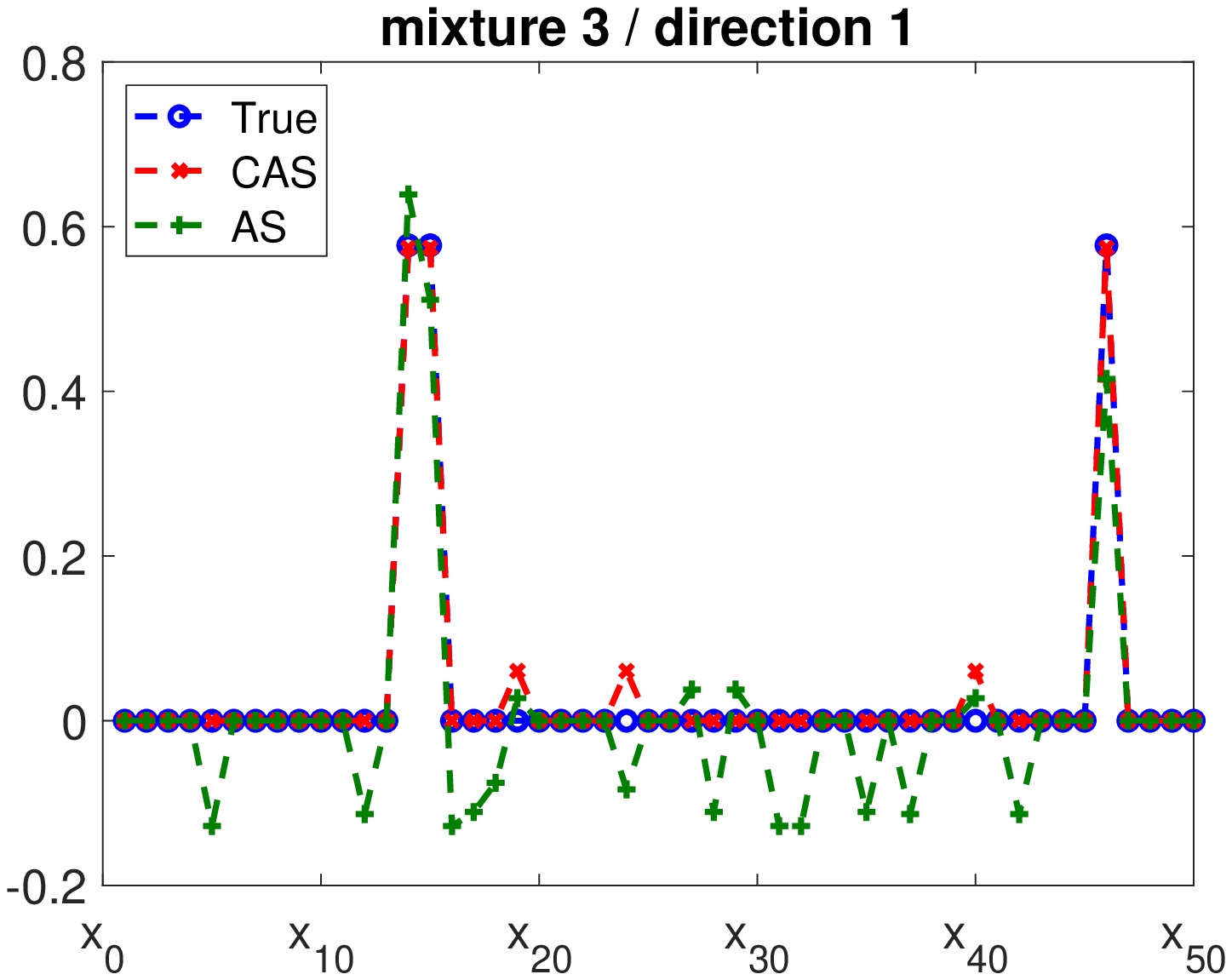}
    \end{minipage}
    \hfill
    \begin{minipage}[t]{0.5\linewidth}
    \small
    \centering
    \includegraphics[width=1\linewidth]{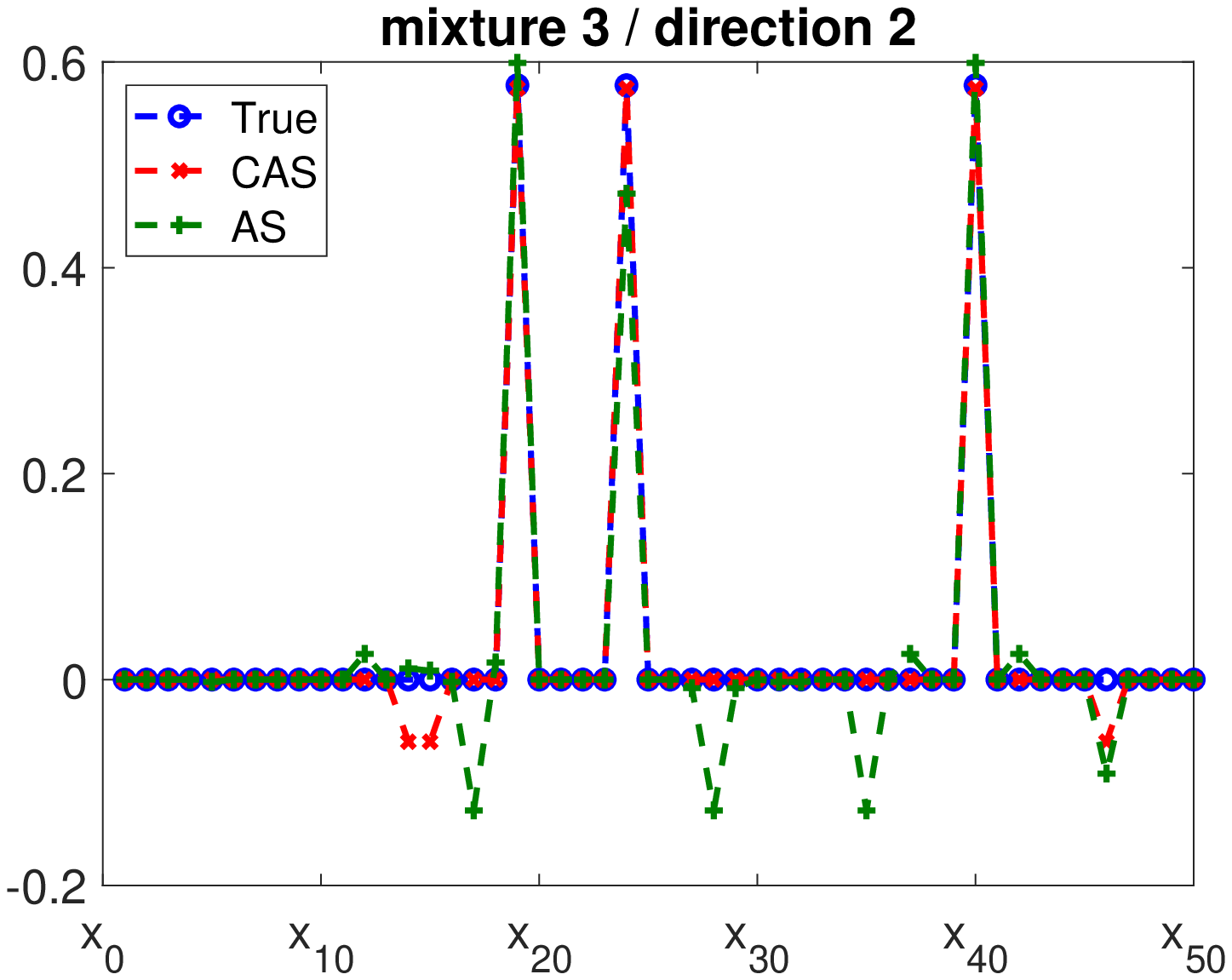}
    \end{minipage}
\end{figure}

The first example is a piece-wise function and it is therefore non-smooth. 
To test the performance with smooth functions, we consider the following mixture of Gaussian functions:
\begin{equation}
  f({\bf x})=\sum_{j=1}^{J}\theta_{j} \exp(-\frac{({\bf x}-{\bf c}_{j})B_{j}B_{j}^{\mathrm{T}}({\bf x}-{\bf c}_{j})^{\mathrm{T}}}{2\sigma_{j}^2})
\end{equation}
 with $B_{j}$ being the low dimensionality projection matrix of size $d\times l$ in which $l\ll d$. 
 Here we set $d=50$, $l=2$ and $J=3$ in this experiment, which means that the mixture has three components, 
  each mixture component admits two intrinsic dimensions and totally six intrinsic dimensions 
exists in this function. To keep it simple and clear, we specify the columns in each $B_{j}$ to be orthogonal
where entries are  randomly drawn according to a binomial distribution such that it is either $0$ or $1$.
It should be clear that the columns of these matrices represent the DR directions in each mixture, 
which are shown in Fig.~\ref{f:dr4gm}.
 Centers ${\bf c}_{j}$ are  also randomly chosen from a uniform distribution $U[0,1]^d$, $\sigma_j$'s are all set to be $0.2$, and the weights are taken to be 
 $\theta=[0.41,\,0.44,\,0.26]$. The distribution of $\-x$ is set to be $U[0,1]^d$.

  In the numerical tests, once again 1000 samples are used as the training set and another 5000 are used as the test set for evaluating the accuracy of GP.  
In Figs.~\ref{f:dr4gm} we plot the actual DR directions as well as the directions computed with CAS and AS,
 where we can see that the two methods can both  identify the DR directions rather accurately. 
  Next we compare the performance  of the GP emulators constructed with the same class of methods used in the first example
  and we summarize all the results in Table~\ref{tb:gm}.
These results are qualitatively similar to those for the first example. 
In particular CAS with 2, 3 or 4 clusters yield substantially lower approximation errors than AS and other methods,
which demonstrates that CAS-LGP can take advantage of the local low dimensional structure 
to achieve better performance than the GP constructed on the global low dimensional subspace,
even though, in both AS and CAS, the global subspace is correctly identified.  
This example shows that the proposed CAS-LGP method performs well in problems where the underlying model does not have a piecewise structure (i.e., the low dimensional structures are not clearly separated in disjoint subdomains). 

    \begin{table}
      \centering
      \caption{The NMSE results of the Gaussian mixture example.}\label{tb:gm}
      \begin{tabular}{|c|c|c|c|c|c|c|c|}
      \hline
      \multicolumn{8}{|c|}{$\eta=1$}\\
      \hline
      Method & \multicolumn{3}{|c|}{CAS-LGP} & \multirow{3}*{AS-GP} & \multirow{3}*{SIR-GP} & \multirow{3}*{SAVE-GP} & \multirow{3}*{GP}\\
      \cline{1-4}     
      \diagbox{d}{k} & 2 & 3 & 4 & & & &\\
      \hline
      1 & 85.7\% & {\bf 85.4}\% &  97.2\%& 89.8\% & 93.2\% & 95.6\% & \multirow{4}*{95.8\%} \\
      \cline{1-7} 
      2 & 25.2\% & {\bf 23.6}\% & 36.6\% & 37.6\% & 92.5\% & 94.8\% &  \\
      \cline{1-7} 
      3 & 24.7\% & {\bf 23.6}\% & 36.6\% & 37.3\% & 92.3\% & 94.3\% &  \\
      \cline{1-7} 
      4 & 24.9\% & {\bf 23.6}\% & 36.6\% & 36.5\% & 92.9\% & 94.6\% &  \\
      \hline
      \multicolumn{8}{|c|}{$\eta=0.5$}\\
      \hline
      Method & \multicolumn{3}{|c|}{CAS-LGP} & \multirow{3}*{AS-GP} & \multirow{3}*{SIR-GP} & \multirow{3}*{SAVE-GP} & \multirow{3}*{GP}\\
      \cline{1-4}     
      \diagbox{d}{k} & 2 & 3 & 4 & & & &\\
      \hline
      1 & 85.7\% & {\bf 85.4}\% &  113.0\% & 89.8\% & 93.2\% & 95.6\% & \multirow{4}*{95.8\%} \\
      \cline{1-7} 
      2 & 25.2\% & {\bf 23.6}\% & 36.7\% & 37.6\% & 92.5\% & 94.8\% &  \\
      \cline{1-7} 
      3 & 24.7\% & {\bf 23.6}\% & 36.6\% & 37.3\% & 92.3\% & 94.3\% &  \\
      \cline{1-7} 
      4 & 24.9\% & {\bf 23.6}\% & 50.1\% & 36.5\% & 92.9\% & 94.6\% &  \\
      \hline
      \end{tabular}
    \end{table}
\subsection{Elliptic PDE}
Our last example is the following elliptic partial differential equation studied in \cite{constantine2014active} with slight modification:
\begin{equation}
  -\nabla_{{\bf s}}\cdot(a(\bf s, \,\bf x)\nabla_{{\bf s}}u)=1,\,\,{\bf s}\in[0,1]^{2}.
\end{equation}
We set homogeneous Dirichlet boundary conditions on the left, top, and bottom of the spatial domain; denote this boundary by $\Gamma_{1}$. The right side of the spatial domain denoted $\Gamma_{2}$ has a homogeneous Neumann boundary condition. That is, 
\begin{equation}
  \begin{aligned}
    u({\bf s})=0,\,\, & {\bf s}\in \Gamma_{1},\\
    \nabla u({\bf s})\cdot {\bf n}=0,\,\, & {\bf s}\in \Gamma_{2}.
  \end{aligned}
\end{equation}
In this problem we assume that the coefficients $a=a({\bf s})$ of the differential operator is a log-Gaussian random field.
Moreover we represent $a(\bf s)$ by a truncated Karhunen-Lo\`{e}ve (KL) type expansion:
\begin{equation}
  \log(a({\bf s}))=\sum_{i=1}^{d}x_{i}{\gamma_{i}}\phi_{i}(\bf s),\label{e:kle}
\end{equation}
where the $x_{i}$ are independent, identically distributed standard normal random variables, and in principle the $\{\phi_{i},\gamma^2_{i}\}$ are 
the eigenpairs of a correlation operator. 
In this example we will modify the standard setting and the eigenpairs $\{\phi_{i},\gamma_{i}^2\}$ will be specified later.
Our function of interest is a linear functional of the solution~\cite{constantine2014active}
\begin{equation}
  f({\bf x})=\int_{\Gamma_{2}}u({\bf s},{\bf x})/|\Gamma_{2}|d{\bf s}.\label{e:foi}
\end{equation}

The PDE is discretized and solved using a finite element method on a triangulation mesh; then $f$ and $\nabla_{{\bf x}}f$ can be computed as a forward and adjoint problem (see \cite{babuska2004galerkin} for details). Recall that by the KL representation~\eqref{e:kle}, we can specify $a({\bf s},{\bf x})$ by providing 
$\{\phi_{i},\gamma_{i}\}$.
First the KL bases $\phi_i$ are taken to be the eigenfunctions of the following covariance kernel function:
\begin{equation}
C(\-{\bf s},\-{\bf s}') = \exp(-\frac{\|\-{\bf s}-\-{\bf s}'\|_1}\beta),
\end{equation}
where $\beta$ is taken to be $1$, 
and $d$ is taken to be 100, implying that the dimensionality of the problem is 100. 
As is mentioned earlier, we modify the 
 eigenvalues $\gamma_{i}$'s
so that they become a function of $\-{\bf x}$. 
Specifically  we assume that the vector-valued function $$\bm\gamma({\bf x})=(\gamma_1(\-{\bf x}),...,\gamma_{100}(\-{\bf x})),$$ takes the following form, 
 \begin{equation}
  \bm{\gamma}({\bf x})=\left\{
  \begin{aligned}
  &\gamma_{3,4,5,6,7,8,9,10}(\-{\bf x})=100 \,\, \& \,\, \gamma_{\Gamma\backslash\{3,4,5,67,8,9,10\}}(\-{\bf x})=0, \quad if \,\, { x}_{1}<0, x_2<0, \\
 &\gamma_{11,12,13,14,15,16,17,18}(\-{\bf x}) =100 \,\, \& \,\,  \gamma_{\Gamma\backslash\{11,12,13,14,15,16,17,18\}}(\-{\bf x})=0, \quad if \,\, { x}_{1}\geq0,{ x}_{2}<0, \\
 &\gamma_{19,20,21,22,23,24,25,26}(\-{\bf x}) =100 \,\, \& \,\,\gamma_{\Gamma\backslash\{19,20,21,22,23,24,25,26\}}(\-{\bf x})=0, \quad if \,\, {x}_{1}<0,{x}_{2}\geq0, \\
 &\gamma_{27,28,29,30,31,32,33,34}(\-{\bf x}) =100 \,\, \& \,\, \gamma_{\Gamma\backslash\{27,28,29,30,31,32,33,34\}}(\-{\bf x})=0, \quad if \,\, 
 {x}_{1}\geq0, {x}_{2}\geq0,
  \end{aligned}\right.
\end{equation}
where $\Gamma$ in this equation means the full index set $\{1,\cdots,100\}$.
Regarding the data, we first generate  random samples from distribution $\pi(\-{\bf x})$,  solve the PDE model, 
and finally evaluate the function of interest~\eqref{e:foi}, yielding the
 input-output pairs, where $1000$ pairs are used as the training set and  $400$ are used  for testing. 
 It is important to mention here that for the 1000 training data, the gradient of the target function is also obtained.

\begin{figure}
    \centerline{\includegraphics[width=.25\linewidth]{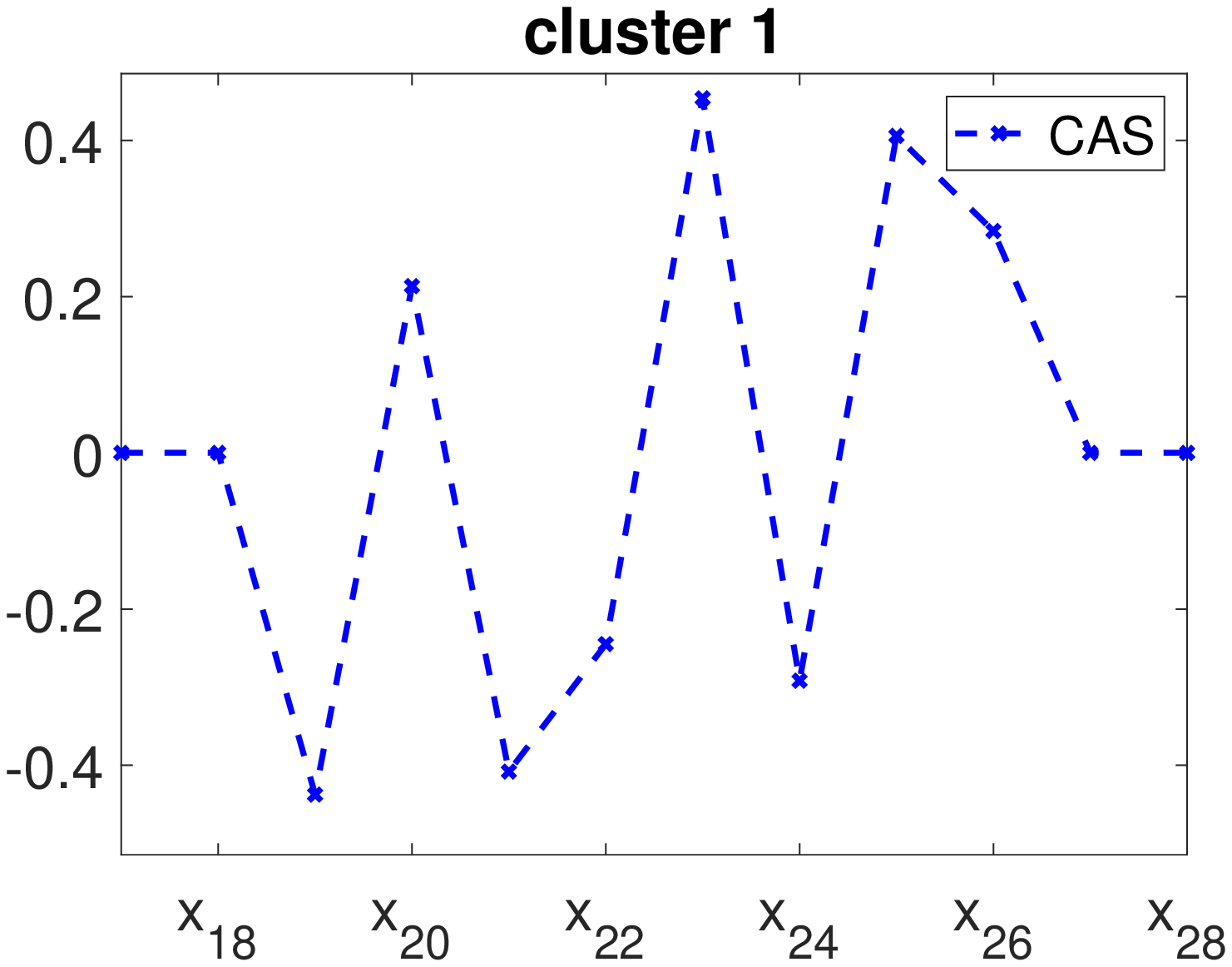}
    \includegraphics[width=.25\linewidth]{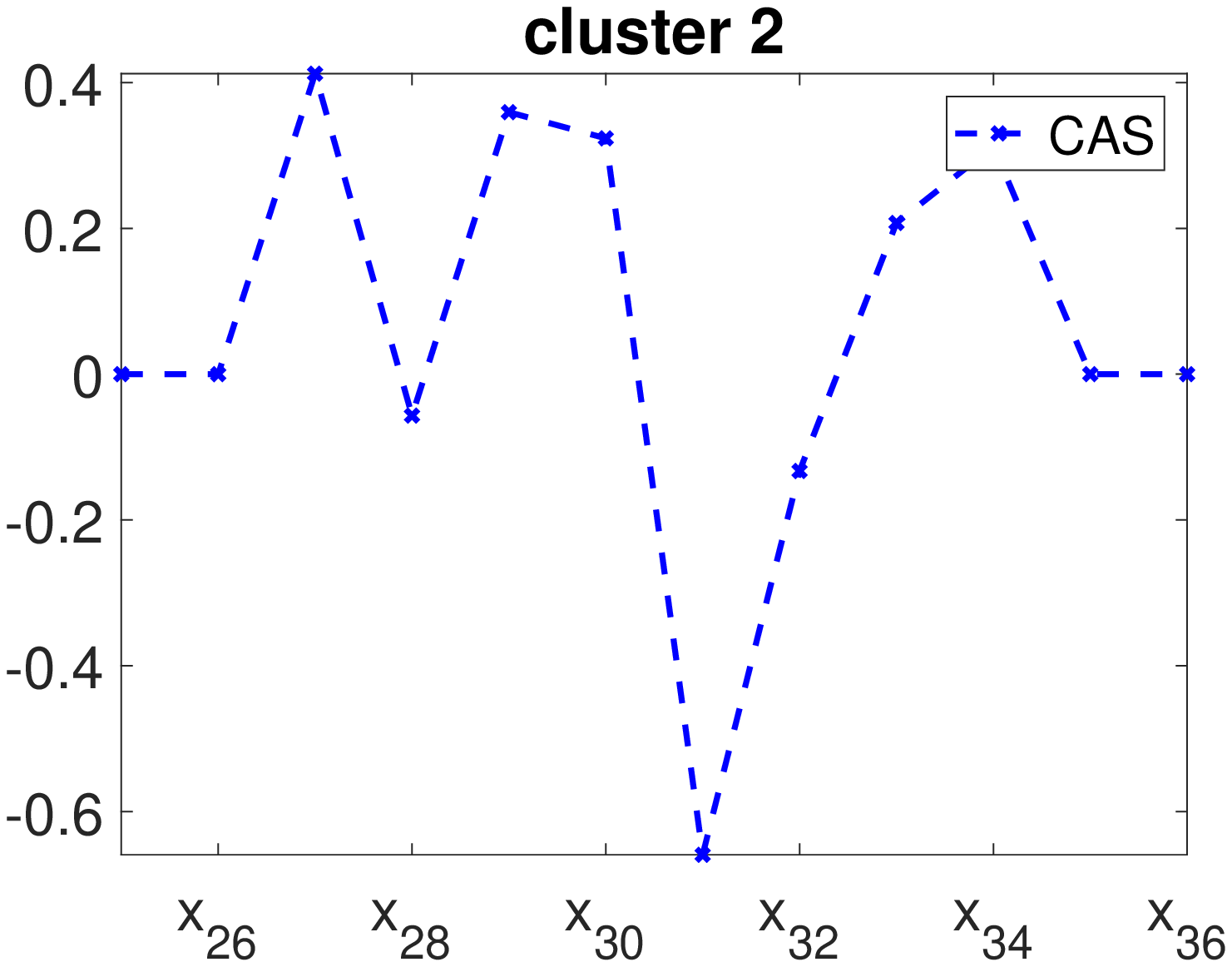}
    \includegraphics[width=.25\linewidth]{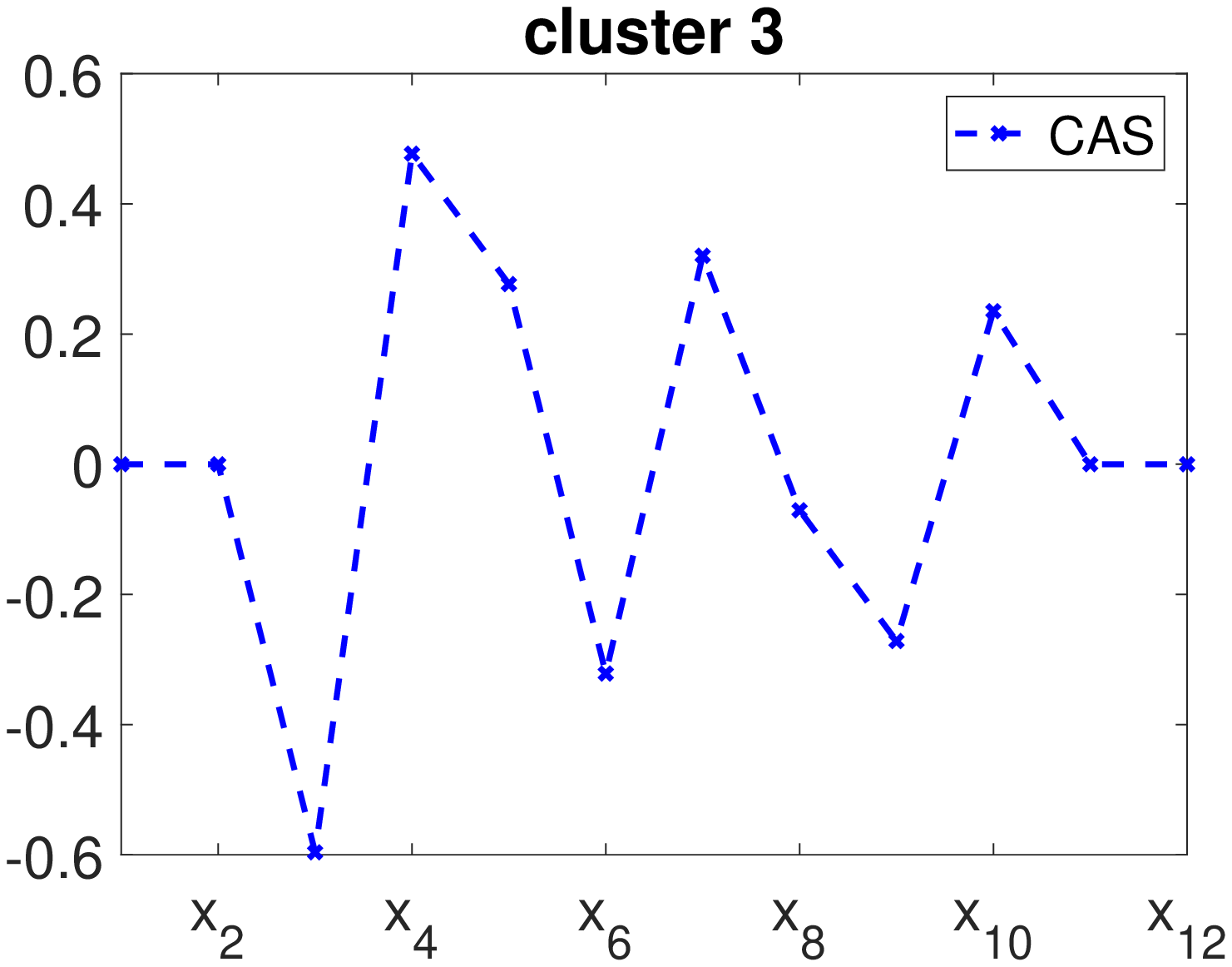}
    \includegraphics[width=.25\linewidth]{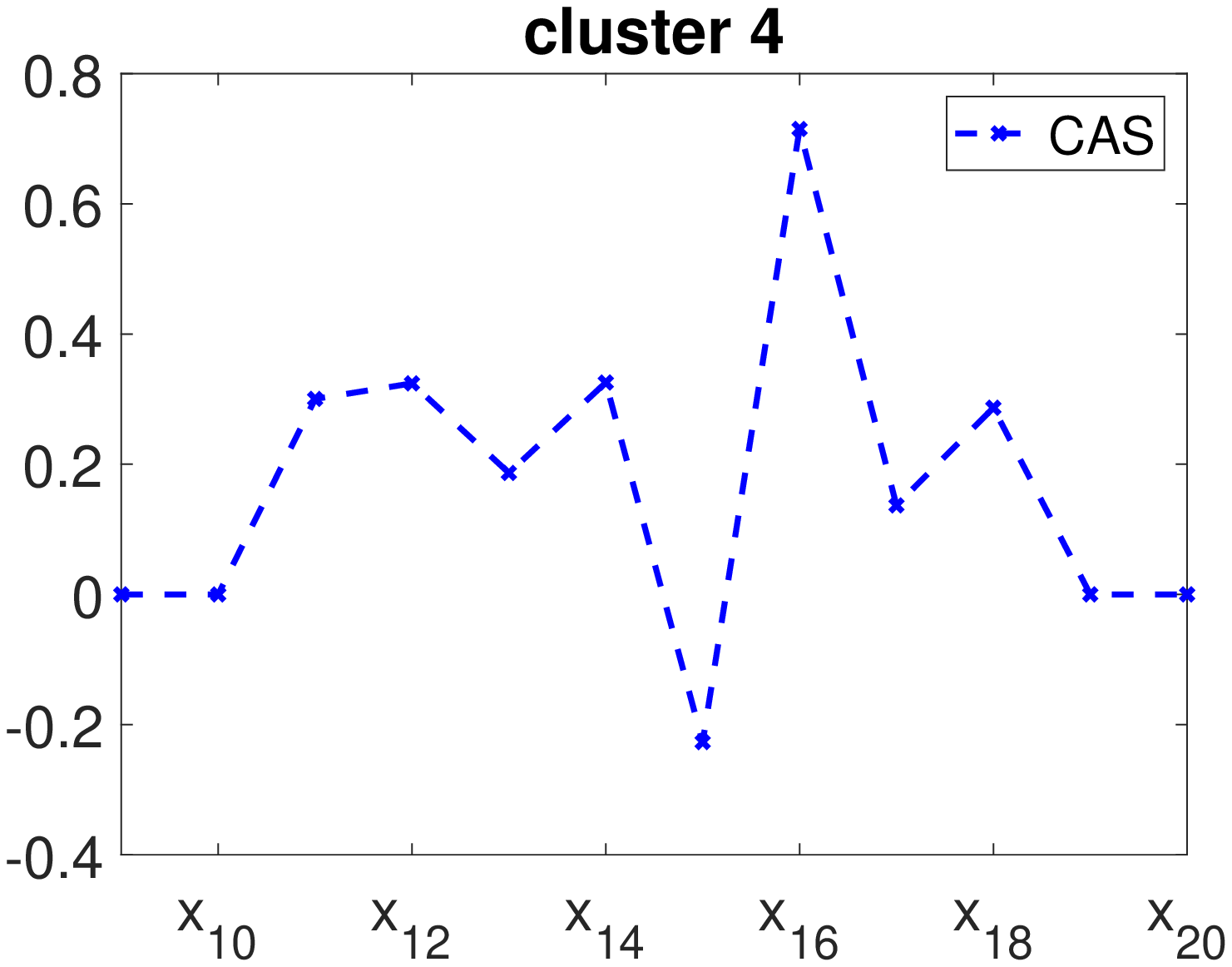}}
      \caption{The leading DR directions in four different clusters calculated by the CAS method. 
      The dimensions that are not shown in the figure are all of zero value.}\label{f:pdedr}
\end{figure}

 First we conduct the comparison of the NMSE results for the same set of methods as those in the first two examples,
 which are shown in Table~\ref{tb:pde}.
  We reinstate that in the case the cluster number is one, CAS reduces to the standard AS.  
  From the table we can see that all the global dimension reduction methods produce comparable results while 
 the performance of CAS is clearly better than those. Within CAS,  the use of four clusters provides the best results.  
 We can also see from the table that with 4 clusters, keeping 1 dimension seems to be sufficient for the local GP emulator and 
 increasing the dimensionality does not improve the performance. 
Moreover, the table also shows that results of CAS are rather robust with respect to the value of $\alpha$. 
Finally for illustration purposes, we plot the leading DR direction in each of the four clusters in Fig.~\ref{f:pdedr}.
  
\begin{figure}
        \centering
        \includegraphics[width=8cm]{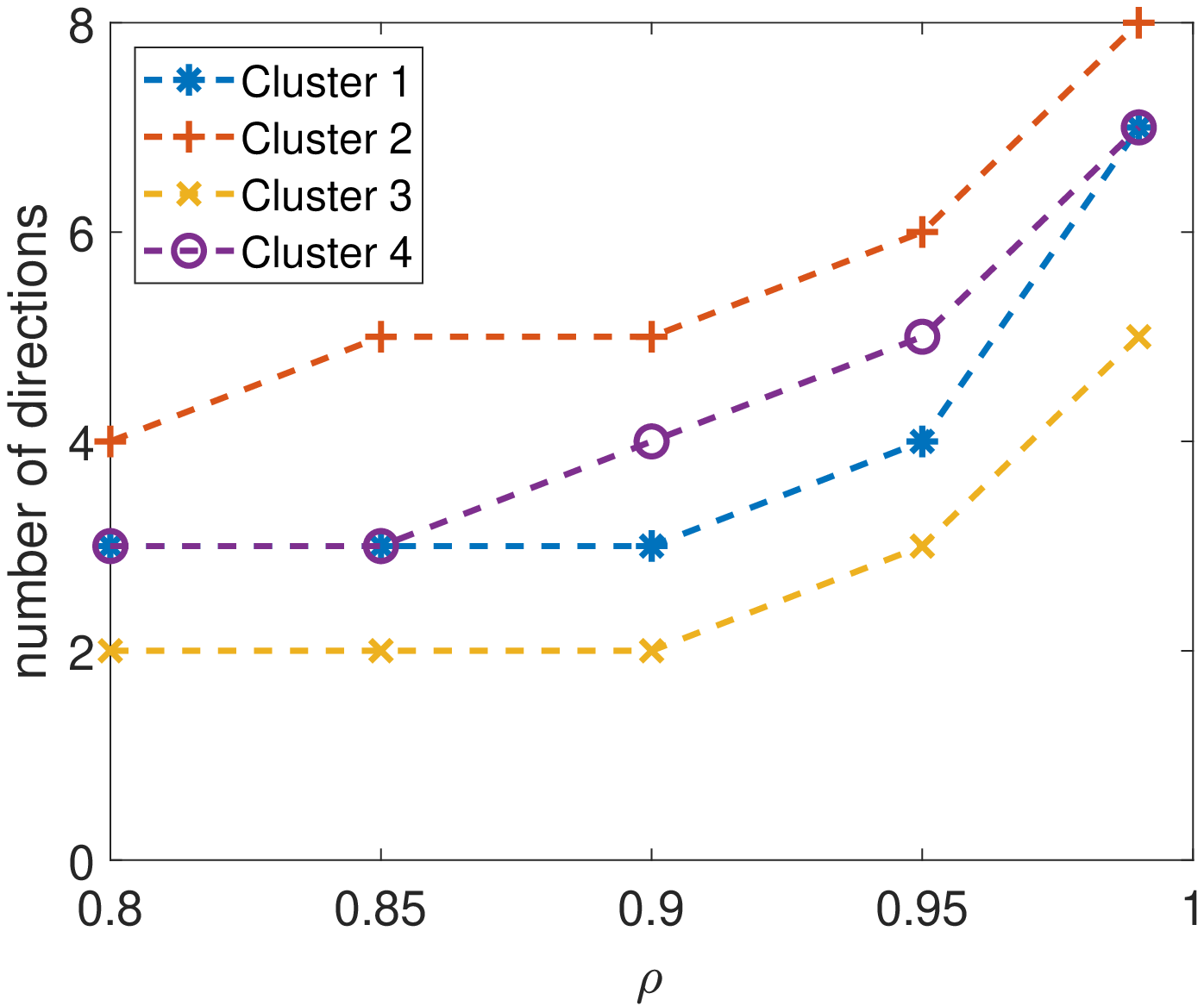}
        \caption{The number of directions determined by the value of $\rho$.}
        \label{fig:nod}
    \end{figure}

Another important issue in CAS method is to determine the number of clusters $J$ and the reduced dimensionality $r_j$ in each cluster. 
In Fig.~\ref{fig:nod} we show the relation between the reduced dimensionality  $r_j$ and the value of $\rho$, where 
one  can see that, while $r_j$ depends on $\rho$, the dependence is not highly sensitive. 
In what follows we conduct the numerical experiments using  Eq.~\eqref{e:rho} with $\rho=0.85$ and 0.95.
The number of clusters is determined by a ten-fold cross validation procedure, the results of which are shown in Table~\ref{tb:cv-pde},
from which we can see that, in both cases the NMSE is decreasing as $J$ increases from 1 to 4 and it remains about the same at $4$ and $5$.  
Based on the  the principle of Occam's razor \cite{rasmussen2001occam} for model selection, we choose $J=4$ here and the final NMSE is $20.4\%$ for $\rho=0.85$
and $21.6\%$ for $\rho=0.95$.
Note that these are the results where
all the $r_j$'s are determined automatically, which is different from those in Table~\ref{tb:pde}.

\begin{table}
      \centering
      \caption{The NMSE results of the PDE example.}\label{tb:pde}
      \begin{tabular}{|c|c|c|c|c|c|c|c|}
      \hline
      \multicolumn{8}{|c|}{$\eta=1$}\\
      \hline
      Method & \multicolumn{3}{|c|}{CAS-LGP} & \multirow{3}*{AS-GP} & \multirow{3}*{SIR-GP} & \multirow{3}*{SAVE-GP} & \multirow{3}*{GP}\\
      \cline{1-4}     
      \diagbox{d}{k} & 2 & 3 & 4 & & & &\\
      \hline
      1 & 27.4\% & 27.3\% & {\bf 20.5\%} & 41.0\% & 37.4\% & 41.0\% & \multirow{4}*{41.0\%} \\
      \cline{1-7} 
      2 & 27.8\% & 27.9\% & {\bf 20.1\%} & 41.0\% & 38.1\% & 41.0\% &  \\
      \cline{1-7} 
      3 & 28.5\% & 28.4\% & {\bf 20.4\%} & 41.0\% & 41.0\% & 41.0\% &  \\
      \cline{1-7} 
      4 & 29.4\% & 29.3\% & {\bf 21.6\%} & 41.0\% & 41.0\% & 41.0\% &  \\
      \hline
      \multicolumn{8}{|c|}{$\eta=0.5$}\\
      \hline
      Method & \multicolumn{3}{|c|}{CAS-LGP} & \multirow{3}*{AS-GP} & \multirow{3}*{SIR-GP} & \multirow{3}*{SAVE-GP} & \multirow{3}*{GP}\\
      \cline{1-4}     
      \diagbox{d}{k} & 2 & 3 & 4 & & & &\\
      \hline
      1 & 29.2\% & 23.1\% & {\bf 20.5\%} & 41.0\% & 37.4\% & 41.0\% & \multirow{4}*{41.0\%} \\
      \cline{1-7} 
      2 & 30.0\% & 23.8\% & {\bf 20.1\%} & 41.0\% & 38.1\% & 41.0\% &  \\
      \cline{1-7} 
      3 & 29.7\% & 24.4\% & {\bf 20.4\%} & 41.0\% & 41.0\% & 41.0\% &  \\
      \cline{1-7} 
      4 & 30.4\% & 25.2\% & {\bf 21.6\%} & 41.0\% & 41.0\% & 41.0\% &  \\
      \hline
      \end{tabular}
    \end{table}
    
\begin{table}
      \centering
      \caption{The Cross Validation results.}\label{tb:cv-pde}
      \begin{tabular}{|c|c|c|c|c|c|c|}
      \hline
      CAS-LGP & \multicolumn{5}{|c|}{CV-NMSE} & NMSE\\
      \hline   
      \diagbox{$\rho$}{J} & 1 & 2 & 3 & {\bf 4} & 5 & 4\\
      \hline
      0.85 & 36.4\% & 32.8\% & 31.3\% & {\bf 30.3\%} & 30.2\% & 20.4\% \\
      \hline
      0.95 & 36.4\% & 36.3\% & 32.4\% & {\bf 30.3\%} & 30.3\% & 21.6\% \\
      \hline
      \end{tabular}
    \end{table}
    
\section{Conclusions}\label{sec:conclu}
In this work, we consider the construction of GP emulators for computationally intensive models of large dimensions.
In such problems, the construction of GP emulator  directly in the original input space is usually 
not feasible due to the fact that GP models can not handle very high dimensionality. 
Thus, a common practice is to first reduce the dimensionality of the original model and then construct 
the GP emulator in the dimension-reduced parameter space. To this end,  the AS method is particularly effective  
as it utilize the gradient information. 
To deal with  models that do not have a simple 
globally low dimensional structure,  we proposed a clustered AS method, which first identifies
the ``clusters'' and then computes local low dimensional structure associated to them. 
Finally a set of local and low dimensional GP emulators are obtained for the underlying simulation model.  
We apply the proposed method to several examples which do not possess a global low dimensional structure, 
and the numerical results show that the CAS based LGP can provide more accurate results than those based on global dimension reduction. 

Several extensions of the proposed method are possible. 
First, one can see that the CAS method does not rely on the GP emulator and can be incorporated with many other surrogates such as polynomial
or RBF based regression models.
To this end a very straightforward extension is to investigate the advantages and potential issues of combining CAS with different surrogate models. 
Second, the performance of the clustering method depends critically on the distance function used. 
In this work we proposed a distance function which combines the gradient and the location information,
and it is certainly interesting to explore other possible distance functions which can further improve the performance. 
 Finally just like the standard AS method, the proposed CAS method also
relies on the gradient information. However, in many real-world applications, the gradient information may 
be unavailable or available with significant error or noise.
To this end, we hope to extend the present CAS approach so that it 
can identify the low-dimensional structure, using the noisy gradient or without using the gradient information at all. 

\section*{Acknowledgments}
This research was supported by the National Natural Science Foundation of China under grant number 11771289.
\appendix
\section{Proof of Theorem~3.2}
\begin{proof}
  First for $j=1,...,J$ we define 
  \begin{equation*}
	f_{j}({\bf x})=\left\{
	\begin{aligned}
		f({\bf x}),&\,\,{\bf x}\in\Omega_{j},\\
		0,&\,\, {\bf x}\notin\Omega_{j},
	\end{aligned}\right.
\end{equation*}
and it should be clear that 
\[f({\bf x})= \sum_{j=1}^{J}f_{j}({\bf x}).\]
Next we have
\begin{equation*}
\begin{aligned}
  \mathbb{E}[(f-\hat{F})^{2}] &= \sum_{j=1}^{J}w_{k}\mathbb{E}_{\pi_{j}}[(f-\hat{F})^{2}] \\
  						&= \sum_{j=1}^{J}w_{j}\mathbb{E}_{\pi_{j}}[(f_{j}-\hat{F}_{j})^{2}] \\
                      &\leq \sum_{j=1}^{J}w_{j}(1+\frac{1}{N_{j}})\alpha'_{j}(\lambda_{r_{j}+1}+\cdots+\lambda_{d}),
\end{aligned}
\end{equation*}
where the last inequality is a direct application of Theorem (2.2). 
Finally letting  $\alpha_j=w_j \alpha'_j$ completes the proof. 
\end{proof}



\section{Cross validation for determining $J$} \label{sec:cv}

Here we describe a $k$-fold cross validation (CV) procedure for determining the number of clusters $J$. 
For each $J=1$ to $J_{\max}$ (i,e., the maximum number of clusters allowed), we perform the following procedure
to calculate the average NMSE associated to $J$: 
\begin{itemize}
\item First one randomly splits the training set $D$ into $k$ equal groups: $D_1$, ..., $D_k$.
\item For $i=1$ to $k$
\begin{itemize}
\item Take  $D_i$ as a CV test set, 
and the rest combined as the CV training set denoted as $D_{-i}$.
\item Apply the CAS-LGP method to the data set $D_{-i}$  and one obtains a local GP emulator, which is then tested on $D_i$,
yielding a NMSE result, denoted as ${NMSE}_i$. 
\end{itemize}
\item Calculate the average NMSE: $NMSE= \frac1k \sum_{i=1}^k NMSE_i$.
\end{itemize}
Finally we choose $J$  that yields the smallest $NMSE$. 

\section{Supplemental result for the PDE example}

In the following table (\ref{tb:beta001}), we provide some the NMSE results of the PDE example with $\beta=0.01$. 
As one can see from the table, in this case, dimension reduction does not improve the regression accuracy at all,
which provides a good example of the limitation of dimension reduction techniques in general. 
      \begin{table}[H]
      \centering
      \caption{The NMSE results of the PDE example.} \label{tb:beta001}
      \begin{tabular}{|c|c|c|c|c|c|c|c|}
      \hline
      \multicolumn{8}{|c|}{$\eta=0.5,\,\beta=0.01$}\\
      \hline
      Method & \multicolumn{3}{|c|}{CAS-LGP} & \multirow{3}*{AS-GP} & \multirow{3}*{SIR-GP} & \multirow{3}*{SAVE-GP} & \multirow{3}*{GP}\\
      \cline{1-4}     
      \diagbox{d}{k} & 2 & 3 & 4 & & & &\\
      \hline
      1 & 12.57\% & 15.39\% & 15.48\% & {\bf 12.44\%} & 15.39\% & 15.39\% & \multirow{4}*{15.39\%} \\
      \cline{1-7} 
      2 & 15.64\% & {\bf 15.39\%} & 15.48\% & 15.42\% & 15.39\% & 15.39\% &  \\
      \cline{1-7} 
      3 & 15.64\% & {\bf 15.39\%} & 15.48\% & 15.42\% & 15.39\% & 15.39\% &  \\
      \cline{1-7} 
      4 & 15.64\% & {\bf 15.39\%} & 15.48\% & 15.42\% & 15.39\% & 15.39\% &  \\
      \hline
      \end{tabular}
    \end{table}
    
\bibliographystyle{plain}
\bibliography{ref}

\end{document}